\documentclass[pra,amsmath,amssymb,twocolumn,longbibliography]{revtex4-2}
\usepackage{graphicx}
\usepackage{dcolumn}
\usepackage{bm}
\usepackage[utf8]{inputenc}
\usepackage[english]{babel}
\usepackage{amsfonts}
\usepackage{hyperref}
\usepackage{physics}
\usepackage{epstopdf}
\usepackage{mathtools}
\usepackage[capitalise]{cleveref}
\usepackage{float}
\usepackage[space]{grffile}
\usepackage{color}
\usepackage[normalem]{ulem}
\usepackage{subfigure}
\usepackage{natbib}
\binoppenalty=\maxdimen
\relpenalty=\maxdimen

\addto\captionsenglish{}
\DeclareMathAlphabet\mathbfcal{OMS}{cmsy}{b}{n}

\def \hrho{\hat{\rho}}
\def \ha{\hat{a}}
\def \nth{\bar{n}_{\rm th}}

\def \hL{\mathbfcal{\hat{L}}}

\def \haL{\boldsymbol{\hat{a}}_{\rm L}}
\def \haR{\boldsymbol{\hat{a}}_{\rm R}}
\def \hone{\boldsymbol{\hat{1}}}

\begin{document}
	
%	\title{Analytic diagonalizing of Lindbladians using continuous weak symmetries}
	\title{Exact solutions of interacting dissipative systems via weak symmetries}
		
	\author{A. McDonald$^{1,2}$ and A. A. Clerk$^1$}
	\affiliation{$^1$Pritzker School of Molecular Engineering, University of Chicago, Chicago, IL 60637, USA\\
	$^2$ Department of Physics, University of Chicago, Chicago, IL 60637, USA}
	
	%%%%%%%%%%%%%%%%%%%%%%%%%%%%%%%%%%%%%%%%%%%%%%%%%%%%%%%%%
	\begin{abstract}
    We demonstrate how the presence of continuous weak symmetry can be used to {\it analytically} diagonalize the Liouvillian of a class Markovian dissipative systems with arbitrary strong interactions or nonlinearity.  This enables an exact description of the full dynamics and dissipative spectrum.  Our method can be viewed as implementing an exact, sector-dependent mean-field decoupling, or alternatively, as a kind of quantum-to-classical mapping.  We focus on two canonical examples: a nonlinear bosonic mode subject to incoherent loss and pumping, and an inhomogeneous quantum Ising model with arbitrary connectivity and local dissipation.  In both cases, we calculate and analyze the full dissipation spectrum.
    Our method is applicable to a variety of other systems, and could provide a powerful new tool for the study of complex driven-dissipative quantum systems.  
	\end{abstract}

	\maketitle
    \textit{Introduction}.---
    Identifying symmetries provides powerful insights into non-dissipative quantum systems, often providing a route towards finding exact descriptions of dynamics and thermal states.  The key ingredient is usually the direct connection between the existence of symmetry and dynamically-conserved quantities.  Turning to dissipative (open) quantum systems, the situation becomes more subtle, as the non-unitary nature of the evolution makes the link between symmetry and conservation laws less direct (see, e.g.~\cite{Lieu_PRL_Symmetry, Victor_Liang_PRA_Symmetries, Buca_Prosen_NJP_2012, Zambrini_Symmetry_PRA_2020, Zhang_JPA_Math_2020, Lieu_Kramers_2021}).  In the typical case of a Markovian system described by a Lindblad master equation, one often has only a so-called ``weak symmetry" \cite{Buca_Prosen_NJP_2012}.  While this symmetry ensures that the generator of the dynamics (i.e.~the Liouvillian) has a block-diagonal structure, it does not guarantee the existence of a true conserved quantity.  Hence, while such weak symmetries can simplify numerical calcuations 
    \cite{Scarlatella_NJP_2019,Marco_Bose_Hubbard_Dimer_NJP_2021}, they are not {\it a priori} a useful tool for obtaining analytic solutions.  
    
    %%%%%%%%%%%
	% FIG S1 Mean Field
	%%%%%%%%%%%
	\begin{figure}[h!]
	\centering
    \includegraphics[width=0.40 \textwidth]{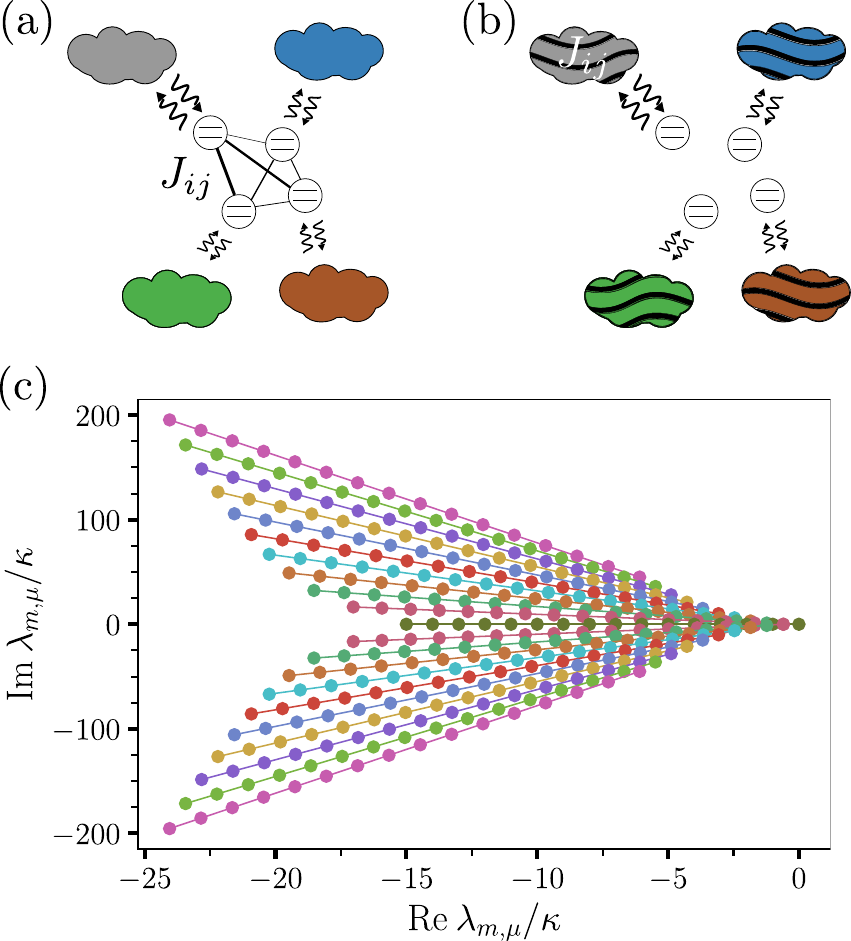}
	\caption{ 
	(a) Schematic of the second model analyzed in this work:  $N$ two-level systems interact via arbitrary Ising interactions $J_{ij}$, and are also subject to local dissipation, c.f.~Eq.~(\ref{eq:SpinMasterEq}).  (b) 
	Using weak symmetry, one can make an {\it exact} mean-field decoupling for each symmetry-constrained dynamical sector, leaving one with an easily-solved but unusual independent, dissipative spin problem.  
    (c) A similar solution method can be used for an 
	incoherently-driven non-linear bosonic mode
	(c.f.~Eq.~(\ref{eq:Kerr_Master_Equation})), enabling an exact calculation of the Liouvillian eigenvalues $\lambda_{m, \mu}$.  We plot these here for $|m| \leq 10$ and $\mu \leq 15$. Each color corresponds to a different value of $|m|$. By fixing $|m|$, the level spacing $\lambda_{m, \mu+1}-\lambda_{m, \mu} = -\tilde{\kappa}_m-i \tilde{U}_m $ is constant, reflecting the non-interacting nature of the problem in each symmetry-constrained block.  We work in a rotating frame where $\omega_0$ is shifted to $0$, and set $U = \kappa, \nth = 0.1$.
	}
	\label{fig:Mean_Field}
	\end{figure}
    %%%%%%%%%%%
    %%%%%%%%%%%%%%%%%%%%%%%%%%%%%%%%%%%%%%%%%%%%%%%%%%%%%%
	%%%%%%%%%%%%%%%%%%%%%%%%%%%%%%%%%%%%%%%%%%%%%%%%%%%%%%
	%%%%%%%%%%%%%%%%%%%%%%%%%%%%%%%%%%%%%%%%%%%%%%%%%%%%%%

    In this Letter, we show that in many cases, the existence of a continuous weak symmetry is in fact a far more powerful tool that one might initially suspect.  We show how weak symmetry can be exploited to fully and \textit{analytically} diagonalize a set of non-trivial Lindblad superoperators that describe interacting, dissipative quantum systems.
    % As we explain in more detail in what follows, this occurs when the generator of the weak-symmetry factorizes in such a way to make mean-field theory exact (see Fig.~\ref{fig:Mean_Field}).  
    As explained below, this is possible because the weak symmetry makes an unusual kind of mean-field decoupling exact in each symmetry-constrained block, reducing it to an effective (but unusual) non-interacting problem (see Fig.~\ref{fig:Mean_Field}).
    Alternatively, the solution method can be viewed as a kind of quantum-to-classical mapping.
    %We demonstrate that when the generator of this weak symmetry factorizes in the correct manner, mean-field becomes exact (see Fig.~\ref{fig:Mean_Field}). 
    The underlying mechanism arises in a wide class of models, but for concreteness, we analyze in detail both a bosonic example (a nonlinear bosonic mode subject to thermal dissipation), and a dissipative spin model (a quantum Ising model subject to single-spin dephasing and relaxation).  Both these examples are directly relevant to a variety of systems under active experimental study.  Our approach yields closed form expressions for all eigenvalues and eigenvectors of the Liouvillian, enabling one to clearly identify structures that would not be apparent otherwise.  This diagonalization provides a full picture of the dissipative dynamics, and also allows the calculation of a variety of observable quantities (e.g.~correlation functions). 
    % The mechanism which makes this possible is generic, but we focus on the thermally-damped non-linear oscillator and a dissipative version of the Ising model. 
     %In the closed version of both models there are conserved quantities, namely the total particle number and $z$-axis magnetization of each spin, which makes the problem trivial. 
%     Although dissipation implies that these are no longer any conserved observables, we will nonetheless show that the remaining weak symmetry still provides a way to analytically tackle the problem. \alexcom{Talk about quantum to classical mapping.}

     We stress that our general method is distinct from approaches used in previous work to obtain exact descriptions of specific quantum dissipative models, e.g.~\cite{Agarwal_Spontaneous_Emission_PRA_1970,  
     Drummond_Exact_P_1980,
     Mark_Old_SPR_1984, Peinov_PRA_Exact_Kerr_Oscillator_1990, Chaturvedi_Exact_Kerr_J_Mod_Opt_1991, Chaturvedi_Exact_Kerr_PRA_1991, 
     Stannigel_2012,
     Torres_Exact_No_Gain_PRA_2014,
     Prosen_Bethe_Ansatz_PRL_2016, 
     Ciuti_Exact_Kerr_PRA_2016,
     Gorkshkov_PRL_2017_Solvable_family_Many-Body,
     Ueda_Bethe_Ansatz_PRL_2018,
     David_PRX_2020,
     Buca_NJP_Bethe_Ansatz_2020, 
     Ueda_PRL_Bethe_Ansatz_2021,
     David_PRX_Quantum_2021}. Our method provides the exact dissipative spectrum and eigenvectors, and moreover, presents them in a simple and intuitive form which is tailor-made to perform analytic computations. This is crucial, as it provides the necessary starting point if one wants to make use of the burgeoning tool of Lindblad perturbation theory \cite{Li2014, LiPRX2016, Ryo_2021_Keldysh_CQED, Victor_Liang_PRX_2106} to more complicated systems. 
     
     %For instance, some models can be solved since the algebraic structure of the Bethe-ansatz is not ruined by dissipation  and in other cases, there is a block diagonal structure which makes the problem computationally efficient . In contrast, our solution method is not simply an extension of a well-known closed-system technique and, further, we can obtain analytical expression for the Lindblad spectrum and eigenvectors.  Studies which compute arbitrary correlation functions using the stochastic wavefunction approach \cite{Plenio_Knight_RMP_1998,Ana_Maria_Diss_Ising_PRA_2013} do not readily have access to this information, a necessary starting point if one wants to make use of the burgeoning tool of Lindblad perturbation theory \cite{Li2014, LiPRX2016, Ryo_2021_Keldysh_CQED, Victor_Liang_PRX_2106} to more complicated systems.
    
    \textit{Dissipative Kerr Oscillator}.---
    Consider a bosonic mode with a Kerr (or Hubbard) type nonlinearity, subject to Markovian thermal dissipation.  The evolution of the system density matrix $\hat{\rho}$ is (setting $\hbar=1$):
    \begin{align} \nonumber
        \partial_t \hrho
        &=
        -i[\omega_0 \ha^\dagger \ha + \frac{U}{2} \ha^\dagger \ha^\dagger \ha \ha
        ,
        \hrho
        ]
        +
        \kappa(\nth+1)\mathcal{D}[\ha] \hrho
        \\ \label{eq:Kerr_Master_Equation}
        &
        +
        \kappa \nth \mathcal{D}[\ha^\dagger]\hrho
        \equiv
        \mathcal{L} \hrho.
    \end{align}
    Here $\ha$ is the mode annihilation operator, $\omega_0$ ($U$) is the mode natural frequency (nonlinearity), $\kappa$ the energy decay rate, and $\nth$ the bath's thermal occupation.  We define $\mathcal{D}[\hat{X}] \hrho \equiv \hat{X}\hrho\hat{X}^\dagger - \{\hat{X}^\dagger \hat{X}, \hrho\}/2$. Eq.~(\ref{eq:Kerr_Master_Equation}) has an obvious weak $U(1)$ symmetry, as it is invariant under $\ha \to e^{-i \theta } \ha$.  This gives $\mathcal{L}$ a block-diagonal structure \cite{Zambrini_Symmetry_PRA_2020, Victor_Liang_PRA_Symmetries, Buca_Prosen_NJP_2012, Marco_Bose_Hubbard_Dimer_NJP_2021, Cristiano_Diss_Phase_Transition_PRA_2018}, which has been used previously to simplify numerical calculations
    \cite{Scarlatella_NJP_2019,Marco_Bose_Hubbard_Dimer_NJP_2021}.  We show below that something more powerful is possible:  despite the nonlinearity, the weak symmetry can also be used to analytically diagonalize \textit{each block} and thus all of $\mathcal{L}$.
    Our analysis complements and extends previous studies that derive exact results for this model without explicit use of weak symmetry  \cite{Peinov_PRA_Exact_Kerr_Oscillator_1990, Mark_Old_SPR_1984, Chaturvedi_Exact_Kerr_J_Mod_Opt_1991, Chaturvedi_Exact_Kerr_PRA_1991}.  In particular, our approach provides simple analytic expressions for {\it all} eigenvalues and eigenvectors of $\mathcal{L}$. 
    
    To diagonalize $\mathcal{L}$, we use the formalism of third-quantization \cite{Prosen_Fermions_2008, Prosen_Bosons_2010, Victor_Thesis}; relevant details can be found in the SM \cite{SM}. %In third-quantization, one thinks of operators as vectors in some Hilbert space $\hrho \to \ket{\hrho}$ and superoperators as operators acting on this new space $\mathcal{L} \to \hL$. The inner product in this Hilbert space is defined by $\langle \hat{B} | \hat{A} \rangle \equiv \Tr \hat{B}^\dagger \hat{A}$, which then allows one to define, in the usual manner, the Hermitian conjugate or adjoint of a superoperator $\hL^\dagger$. 
    % To obtain a compact representation of $\mathcal{L}$, 
    One first introduces four new superoperators $\haL \ket{\hrho} \equiv \ket{\ha \hrho}$, $\haR \ket{\hrho} \equiv | \hrho \ha \rangle$, $\haL^\dagger \ket{\hrho} \equiv |\ha^\dagger \hrho \rangle$, and $\haR^\dagger \ket{\hrho} \equiv |\hrho \ha^\dagger\rangle$
   %\begin{align}\label{eq:Def_L_R_a}
	%\haL \ket{\hrho} \equiv \ket{\ha \hrho},
	%\hspace{0.5cm}
	%&\haR \ket{\hrho} \equiv | \hrho \ha \rangle
	%\\ \label{eq:Def_L_R_a_dag}
	%\haL^\dagger \ket{\hrho} \equiv |\ha^\dagger \hrho \rangle,
	%\hspace{0.4cm}
	%&\haR^\dagger \ket{\hrho} \equiv |\hrho \ha^\dagger\rangle
    %\end{align}
    which we will refer to as annihilation and creation superoperators. We will also reserve the bold typeface to indicate a third-quantized superoperator $\mathcal{L} \to \hL$. %Note that $\boldsymbol{\ha}_{\rm L/R}^\dagger$ are in fact the Hermitian conjugate of $\boldsymbol{\ha}_{\rm L/R}$, which can be easily shown using the cyclic invariance of the trace. 
    % If we subtract for the moment a trivial contribution to the Lindbladian $(i \omega_0+\kappa/2)\hone$, with $\hone$ the identity superoperator, then we
    We can now express our Liouvillian as 
    $\hL = (i \omega_0+\kappa/2)\hone + \hL_0 + \hL_{\rm int}$ where
    \begin{align}
        \hL_0 
        &=
        \underline{\boldsymbol{\hat{a}}}^\dagger
        \begin{pmatrix}
        -i \omega_0-\frac{\kappa}{2}(2\nth+1) & \kappa \nth 
        \\
        \kappa(\nth+1) & i\omega_0-\frac{\kappa}{2}(2\nth+1)
        \end{pmatrix}
        \underline{\boldsymbol{\hat{a}}}
        \\ \label{eq:L_Thermal_Kerr}
        \hL_{\rm int}
        &=
        -i
        \frac{U}{2}
        \left( 
        \haL^\dagger \haL-\haR \haR^\dagger
        \right)
        \left(
        \haL^\dagger \haL+\haR \haR^\dagger
        -\hone
        \right)
    \end{align}
    %\begin{align}\nonumber
    %    &\hL
    %    =
    %    \hL_0+\hL_{\rm int}
    %    \\ \nonumber
    %    &=
    %    \underline{\boldsymbol{\hat{a}}}^\dagger
    %    \begin{pmatrix}
    %    -i \omega_0-\frac{\kappa}{2}(2\nth+1) & \kappa \nth 
    %    \\
    %    \kappa(\nth+1) & i\omega_0-\frac{\kappa}{2}(2\nth+1)
    %    \end{pmatrix}
    %    \underline{\boldsymbol{\hat{a}}}
        %\\ \nonumber
        %&
        %+(-i\omega_0+ \frac{\kappa}{2}) \hone
    %       \\ \label{eq:L_Thermal_Kerr}
    %    &-i
    %    \frac{U}{2}
    %    \left( 
    %    \haL^\dagger \haL-\haR \haR^\dagger
    %    \right)
    %    \left(
    %    \haL^\dagger \haL+\haR \haR^\dagger
    %    -\hone
    %    \right)
    %\end{align}
    %\begin{align} \nonumber
    %    &\hL 
    %    =
    %    (-i\omega_0-\frac{\kappa}{2}(2\nth+1))\haL^\dagger \haL
    %    +
    %    \kappa(\nth+1)\haL \haR^\dagger
    %    \\ \nonumber
    %    &+
    %    (i \omega_0-\frac{\kappa}{2}(2\nth+1))\haR \haR^\dagger
    %    + \kappa \nth \haL^\dagger \haR
    %    - \kappa \nth \hone
    %    \\ \label{eq:L_Thermal_Kerr}
    %    &
    %    -i
    %    \frac{U}{2}
    %    \left( 
    %    \haL^\dagger \haL-\haR \haR^\dagger
    %    \right)
    %    \left(
    %    \haL^\dagger \haL+\haR \haR^\dagger
    %    -\hone
    %    \right)
    %\end{align}
    correspond to the quadratic and interacting parts of the Linbladian respectively. Here $\underline{\boldsymbol{\hat{a}}}^\dagger = \begin{pmatrix}
    \haL^\dagger & \haR^\dagger
    \end{pmatrix}$.
    % is a row of superoperators. 
    %and we have used $[\haL, \haL^\dagger] = -[\haR, \haR^\dagger] = \hone$ along with the fact that $\haL, \haL^\dagger$ commute with $\haR, \haR^\dagger$ since they act on different sides of the density matrix. Note that we have factorized the interaction in a manner which will soon prove to be useful.
    The quadratic part of the superoperator $\hL_0$ is easily diagonalized via standard third-quantization techniques \cite{Prosen_Fermions_2008, Prosen_Bosons_2010}.
    % \alexcom{Could save space by directly stating that $\hL_{\rm int}$ is where the difficulty lies and thus not even writing $\mathcal{L}_0$ explicitly.}. 
    The nonlinear quartic terms however represent a true interaction of third-quantized bosons, and at first glance destroy exact solvability.

	We now exploit the weak symmetry of our system. At the superoperator level, the weak symmetry corresponds to the invariance of Eq.~(\ref{eq:L_Thermal_Kerr}) under $\boldsymbol{\hat{a}}_{\rm L/R} \to \boldsymbol{\hat{a}}_{\rm L/R} e^{-i \theta}$.
	%In this case there is an obvious $U(1)$ symmetry, since the master equation Eq.~(\ref{eq:Kerr_Master_Equation}) remains unchanged under an arbitrary change of phase of the mode annihilation operator $\ha \to e^{-i \theta} \ha$. At the superoperator level, this is equivalent to the statement that the third-quantized Lindbladian in Eq.~(\ref{eq:L_Thermal_Kerr}) is invariant under $\boldsymbol{\hat{a}}_{\rm L/R} \to \boldsymbol{\hat{a}}_{\rm L/R} e^{-i \theta}$.
	The superoperator generating this effective unitary transformation is $\haL^\dagger \haL-\haR \haR^\dagger$, which immediately implies $[\hL, \haL^\dagger \haL-\haR \haR^\dagger] = 0$.
	%Using the superoperator commutation relations, one can show that the associated unitary operator which implements this transformation is $e^{i \theta()}  \boldsymbol{\hat{a}}_{\rm L/R} e^{-i \theta(\haL^\dagger \haL-\haR \haR^\dagger)} = e^{-i \theta} \boldsymbol{\hat{a}}_{\rm L/R} $. It follows that the Lindbladian superoperator commutes with the Hermitian generator of this $U(1)$ symmetry $[\hL, \haL^\dagger \haL-\haR \haR^\dagger] = 0$. 
	Standard linear algebra then dictates that $\hL$ is block-diagonal in the eigenbasis of $\haL^\dagger \haL-\haR \haR^\dagger$. We can thus write $\hL = \bigoplus_m \hL_m$, where each block $\hL_m$ is indexed by $m$, an eigenvalue of $\boldsymbol{\hat{m}} \equiv \haL^\dagger \haL-\haR \haR^\dagger$. A simple calculation reveals that any outer-product of Fock states $\ket{p}\bra{q}$ is an eigenvector of the generator $\boldsymbol{\hat{m}} \ket{p}\bra{q} =  [\hat{a}^\dagger \hat{a}, \ket{p}\bra{q}] = m \ket{p}\bra{q}$ and the corresponding eigenvalue $m = p-q \in \mathbb{Z}$ characterizes the degree of coherence or off-diagonalness in Fock space. Further, since any outer product of Fock states of the form $\ket{p+n}\bra{q+n}$ has the same eigenvalue as $\ket{p}\bra{q}$, each block $\hL_m$ is infinite in extent.
	
	%Using a weak symmetry to make a Lindblad superoperator block-diagonal is a well-known result \cite{Zambrini_Symmetry_PRA_2020, Victor_Liang_PRA_Symmetries, Buca_Prosen_NJP_2012} which has for example been used to obtain a faster way to numerically solve the master equation \cite{Marco_Bose_Hubbard_Dimer_NJP_2021} \alexcom{Other works that do this explicitly?} in addition to being the crucial first step when studying symmetry-breaking dissipative phase transitions \cite{Cristiano_Diss_Phase_Transition_PRA_2018} . 

    While weak symmetry provides a block-diagonal structure, we are still left with the seemingly formidable task of diagonalizing the infinite-dimensional matrix corresponding to each block; further, apart from $m=0$, each block's form depends on the non-trivial interaction $U$.  As we now show, surprisingly these remaining tasks can be done exactly.    
    % Given that our goal is to analytically diagonalize $\hL$, it seems as though identifying the $U(1)$ symmetry was a pointless endeavor; we must still find the eigenvalues and eigenvectors of infinitely-many infinite-sized matrices, each with a quartic non-linearity. This initial assessment however is incorrect. 
    By definition $\hL_m$, is the full Lindbladian projected onto the subspace spanned by eigenvectors of $\boldsymbol{\hat{m}}$ with eigenvalue $m$. We may thus, in each block $\hL_m$, make the substitution $\boldsymbol{\hat{m}} \to m$. Next, note that the non-linear part of $\mathcal{L}$ can be written as
	\begin{align}\label{eq:Factorize}
	    \hL_{\rm int}
	    =
	    -i \frac{U}{2}
	    \boldsymbol{\hat{m}}
	    \times
	    \hL_0'
	\end{align}
	where $\hL_0' = \left( \haL^\dagger \haL+\haR \haR^\dagger-\hone \right)$ is quadratic in creation an annihilation superoperators. Projecting onto the subspace indexed by $m$, when have $\hL_{\rm int} \to -iUm/2 \hL_0'$. %Reinstating the constant shift  shift $(i \omega_0 + \kappa/2)\hone$ to $\hL$ 
	We finally obtain
	\begin{align}\nonumber
	\hL_m 
	&=
	\underline{\boldsymbol{\hat{a}}}^\dagger
    \begin{pmatrix}
    -i\frac{U m}{2}-\frac{\kappa}{2}(2\nth+1) & \kappa \nth 
    \\
    \kappa(\nth+1) & -i\frac{U m}{2}-\frac{\kappa}{2}(2\nth+1)
    \end{pmatrix}
    \underline{\boldsymbol{\hat{a}}}
	\\ \label{eq:L_m}
	&+
	(
	- i (\omega_0-U)m 
	+
	\frac{\kappa}{2}
	)\hone.
	\end{align}
	%\begin{align} \nonumber
	%    &\hL_m
	%    =
	%    -i(\omega_0-\frac{U}{2})m \hone
	%    +
	%    \kappa(\nth+1)\haL \haR^\dagger
	%    + 
	%    \kappa \nth \haL^\dagger \haR 
	%    \\ \label{eq:L_m}
	%    &
	%    +
	%    (-i \frac{U}{2}m- \frac{\kappa}{2}(2\nth+1))
	%    \left(\haL^\dagger \haL+
	%    \haR
	%    \haR^\dagger \right) 
	%    -
	%    \kappa \nth \hone. 
	%\end{align}
	
	We thus have a crucial first result:  in each symmetry-constrained sector, $\hL$ becomes quadratic in creation and annihilation superoperators, and can thus be diagonalized exactly.  It is as though a mean-field ansatz has become exact in each block (though note the mean-field decoupling is block dependent, and results in a Liouvillian that is not in Lindblad form).  
	We stress that the mere existence of a weak symmetry was not enough for solvability, as this by itself only guarantees the existence of the block-diagonal structure. Instead, we also needed the interacting part of the Lindbladian to factor as in Eq.~(\ref{eq:Factorize}).
% 	, such that once we project onto each block the generator is replaced by a number, whence the problem becomes tractable. 
    Identifying this general structure is a main result of this work. 
    
    As it is quadratic in creation and annihilation superoperators,  Eq.~(\ref{eq:L_m}) can be diagonalized using conventional third-quantization. One ultimately needs to diagonalize a $2 \times 2$ matrix in each sector to obtain both the eigenvalues and eigenvectors. 
    % \alexdelete{Let us note here however that Eq.~(\ref{eq:L_m}) immediately implies that the eigenvectors do not depend on the oscillator's resonant frequency $\omega_0$, unlike the non-linearity $U$ which explicitly appears on the diagonal of the relevant matrix.}
	We denote the Liouvillian eigenvalues $\lambda_{m,\mu}$ where $m$ labels the different symmetry-constrained blocks (i.e.~the degree of off-diagonalness), and the non-negative integer $\mu$ labels eigenmodes in a given block.  It roughly characterizes the average number of particles in the eigenmode. 
	% These eigenvalues determine the characteristic decay rates and oscillation frequencies that characterize the system's dynamics.  
	Using the above structure (see SM \cite{SM}), we find:
	%The right and left eigenvectors of $\hL$ are given by
	%\begin{align}
	%   \ket{\hat{r}_{m,\mu}}
	%    &=
	%    C_{m,\mu}
	%    \begin{cases}
	%    (\hcp^\dagger \hcm)^\mu(\hcp^\dagger)^m
	%    \ket{\hat{0}_m^r}
	%    ,
	%    &
	%    m \geq 0
	%    \\
	%    (\hcp^\dagger \hcm)^\mu(-\hcm^\dagger)^{-m}
	%    \ket{\hat{0}_m^r}
	%    ,
	%    &
	%    m < 0
	%    \end{cases}
	%    \\
	%   \langle \hat{l}_{m,\mu} |
	%    &=
    %   C_{m,\mu}
	%    \begin{cases}
	%    \bra{\hat{0}_m^l}
	%    (\hdp)^m(-\hdm^\dagger \hdp)^\mu
	%    ,
	%    &
	%    m \geq 0
	%    \\
	%    \bra{\hat{0}_m^l}
	%    (\hdm^\dagger)^{-m}(-\hdm^\dagger \hdp)^\mu
	%    ,
	%    &
	%    m < 0
	%    \end{cases}
	%\end{align}
	%where $\mu$ is a non-negative integer labelling the eigenstate, and $C_{m,\mu} = (\mu!(\mu+|m|)!)^{-1/2}$ is a normalization constant. The superoperators $\hcm$,$\hdp$ and $\hcp^\dagger, \hdm^\dagger$ are linear combinations of $\haL, \haR$ and $\haL^\dagger, \haR^\dagger$ respectively. The coeffecients relating these superoperators are defined such that they satisfy a $2 \times 2$ matrix eigenvalue problem. Further, the only non-vanishing commutators between these superoperators are $[\hdp, \hcp^\dagger] = [ \hcm, \hdm^\dagger] = \hone$. The Gaussian operators $\hat{0}^r_m$ and $\hat{0}^l_m$ satisfy $\hdp |\hat{0}^r_m\rangle = \hdm^\dagger |\hat{0}^r_m\rangle = \langle \hat{0}^l_m | \hcm = \langle \hat{0}^l_m | \hcp^\dagger = 0  $.  Finally, we have chosen the normalization such that the biorthogonality condition $\langle \hat{l}_{m,\mu} | \hat{r}_{m',\mu'} \rangle = \delta_{m ,m'} \delta_{\mu,\mu'}$ holds. 
	\begin{align} \nonumber
	    \lambda_{m,\mu}
	    &=
	    -i \left[\omega_0-U
	    +\frac{\tilde{U}_m}{2}(|m|+1+2\mu)
	    \right] m
	    \\ \label{eq:Eigenvalues}
	    &-\frac{1}{2}
	    \Big[
	    \tilde{\kappa}_m(|m|+1+2\mu)-\kappa
	    \Big]
	\end{align}
	where 
	\begin{align} \label{eq:U_m}
	    \tilde{U}_m
	    &=
	    |U| \: 
	    \text{Im}
	    \sqrt{(\frac{\kappa}{U m})^2-1+2 i \frac{\kappa}{U m} (2\nth+1)}
	    \\ \label{eq:kappa_m}
	    \tilde{\kappa}_m
	    &=
	    \kappa \:
	    \text{Re}
	    \sqrt{1-(\frac{U m}{\kappa})^2 + 2 i \frac{U m}{\kappa} (2\nth+1)}
	\end{align}
	are renormalized sector-dependent non-linearities and decay rates respectively.  If $\kappa \rightarrow 0$, $\tilde{U}_m \rightarrow U$, whereas for non-zero $\kappa$ it is temperature dependent. 
	We also see that the effective damping rate in each sector generically depends on temperature when $U \neq 0$.
% 	As expected, modes with larger coherences $m$ oscillate and decay more quickly. 
% 	Further, modes with a larger $\mu$ also decay more rapidly, but only enter the oscillation rate through the non-linearity, just as for an undamped non-linear oscillator. 
    In Fig.~\ref{fig:Mean_Field} we plot the spectrum for $|m| \leq 10$ and $\mu  \leq 15$.
	Expressions for eigenvectors are provided in the SM \cite{SM}.

	%%%%%%%%%%%
	% FIG S2 Spectrum
	%%%%%%%%%%%
	%\begin{figure}[t]
	%\centering
    %\includegraphics[width=0.45 \textwidth]{Weak-Symmetries/Main/Spectrum.pdf}
	%\caption{ \alexnew{Eigenvalues $\lambda_{m,\mu}$ of the Lindbladian describing a thermally-damped non-linear oscillator for $|m| \leq 10$ and $\mu \leq 15$. Each color corresponds to a different value of $|m|$; the larger the slope the larger the value of $|m|$. The salient feature is that by fixing $m$, the level spacing $\lambda_{m,\mu+1}-\lambda_{m, \mu} = -\tilde{\kappa}_m-i \tilde{U}_m $ is constant. This simply reflects the effective  quadratic nature of the problem in each block labeled by $m$. Parameters: $\omega_0 = 1, \kappa = U = 0.2, \nth = 0.1$.}
	%\AC{Thought: combine Fig. 1 and 2?}
	%}
	%\label{fig:Spectrum}
	%\end{figure}
    %%%%%%%%%%%
    %%%%%%%%%%%%%%%%%%%%%%%%%%%%%%%%%%%%%%%%%%%%%%%%%%%%%%
	%%%%%%%%%%%%%%%%%%%%%%%%%%%%%%%%%%%%%%%%%%%%%%%%%%%%%%
	%%%%%%%%%%%%%%%%%%%%%%%%%%%%%%%%%%%%%%%%%%%%%%%%%%%%%%
	
	The ability to analytically describe the eiegenvectors and eigenvalues evidently constitutes a full solution of our system: any quantity we wish to calculate or initial state we wish to time-evolve can be readily computed using the spectral decomposition of $\hL$. 
% 	To the best of our knowledge, the exact eigenvalues and eigenvectors had not been previously described in the literature. 
    This spectral information in and of itself carries a wealth of physically and experimentally relevant information. We will focus on two such examples. It is for instance interesting to note that $\tilde{U}_m$ and $\tilde{\kappa}_{m}$ obey 
	\begin{align}
	    &|U| \leq |\tilde{U}_m| < |U|(2\nth+1),
	    \\
	    &\kappa \leq \tilde{\kappa}_m < \kappa(2\nth+1),
	\end{align}
	where the lower bound in both cases is reached if and only if $\nth = 0$. For a fixed non-zero temperature, the dimensionless parameter $\kappa/(U m)$ determines how close one is to reaching the lower or upper bound. For strong non-linearity or large coherences $\kappa \ll U m$ we obtain $\tilde{\kappa}_m \lesssim \kappa(2\nth+1)$. As explained in Sec.~I of the \cite{SM}, in this limit the right and left eigenvectors are simply outer product of Fock states and the real part of $\lambda_{m,\mu}$ corresponds to the average of the Fermi's Golden rule decay rate for the Fock states $\ket{\mu+ |m|}$ and $\ket{\mu}$ \cite{Mark_Old_SPR_1984}. We plot $\tilde{\kappa}_m$ as a function of $m$ for different values of $U$ in Fig.~S1 of the SM \cite{SM}. In the opposite limit of strong dissipation $\kappa \gg U m$ we see that the non-linearity scales linearly with temperature $\tilde{U}_m \lesssim U(2\nth+1)$. If we were to probe a single-particle quantity like the retarded Green's function $G^R(t) =	-i \Theta(t) \langle [ \ha(t), \ha^\dagger(0)] \rangle $, this corresponds to the Hartree-type frequency shift one would obtain via leading-order perturbation theory.
	
	%We first expand both to first order in $\nth$, from which we get
	%\begin{align}\label{eq:U_m_small_nth}
	%    \tilde{U}_m
	%    =
	%    U
	%    +
	%   \frac{2 U \kappa^2}{U^2 m^2+\kappa^2} \nth
	%   +
	%   \mathcal{O}(\nth^2).
	%   \\ \label{eq:kappa_m_small_nth}
	%   \tilde{\kappa}_m
	%   =
	%   \kappa
	%   +
	%   \frac{2 U^2 m^2 \kappa }{U^2 %m^2+\kappa^2}\nth
	%   +
	%   \mathcal{O}(\nth^2).
	%\end{align}
	%We have thus obtained a non-trivial result using the exact analytic treatment: vacuum fluctuations do not renormalize the non-linearity or decay rate of a dissipative Kerr oscillator. In limit where $U \gg \kappa$, Fock states are well resolved and Eq.~(\ref{eq:kappa_m_small_nth}) recovers to Fermi's golde rule $\tilde{\kappa}_m \approx \kappa(2\nth+1)$. In opposite limit of strong dissipation where $\kappa \gg U m$ we see that the non-linearity scales linearly with temperature $\tilde{U}_m \approx U(2\nth+1)$. If we were to probe a single-particle quantity like the retarded Green's function $G^R(t) =	-i \Theta(t) \langle [ \ha(t), \ha^\dagger(0)] \rangle $, this corresponds to the Hartree shift one would see in first-order perturbation theory. \alexcom{I had something here about the $\sqrt{\nth}$ scaling for large $\nth$, but if we can't explain it, perhaps its not to discuss it...}

	The retarded Green's function, which controls how the average value $\langle \ha(t) \rangle$ changes in response to a weak coherent drive applied at time $t = 0$, can of course be computed to all orders in $U$. 
	%Regardless of the parameter regime we are in, it is easy to demonstrate that any amount of thermal quanta $\nth$ increases both oscillation and decay rates. Finally, we note that in the limit of large temperature we have $\tilde{U}_m \sim \sqrt{\nth/m}$ and $\tilde{\kappa}_m \sim \sqrt{\nth}$.\alexcom{Again, still haven't been able to find a suitable reason why the scaling is like $\sqrt{\nth}$ for large temperature...should we even keep this?}
	%The ability to analytically find the eigenvectors and eigenvalues evidently constitutes a full solution to the problem at hand; any quantity we wish to calculate can be readily computed using the spectral decomposition of $\hL$. 
     Since $\hrho_{\rm ss}$ is an incoherent mixture of Fock states, it is an element of the $m = 0$ block. Applying $\ha^\dagger$ to either side of the density matrix raises the coherence by one, and thus excites all $m = 1$ right eigenvectors. Using the spectral decomposition of $e^{\hL t}$ we show in the SM \cite{SM} that
	%\begin{align}
	%    &G^R(t)=
	%    -i \Theta(t)
	%    \frac
	%    {e^{-i(\omega_0-U)t+ \frac{\kappa}{2}t}}
	%      {
    %    \left(
    %    \cosh (\frac{\tilde{\kappa}_1+i %\tilde{U}_1}{2}t)
    %    +
    %    R_1
    %    \sinh (\frac{\tilde{\kappa}_1+i \tilde{U}_1}{2}t)
    %    \right)^2}
	%\end{align}
	\begin{align} \nonumber
	&G^R(t)
	=
	    -i \Theta(t)
	    \sum_{\mu = 0}^\infty
	    e^{\lambda_{m=1, \mu} t}
	    \Tr(\ha \hat{r}_{m=1, \mu})
	    \Tr(\hat{l}^\dagger_{m=1, \mu}[\ha^\dagger, \hrho_{\rm ss}])
	    \\ \label{eq:GR}
	   &=
	    -i \Theta(t)
	    \frac
	    {e^{-i(\omega_0-U)t+ \frac{\kappa}{2}t}}
	      {
        \left(
        \cosh (\frac{\tilde{\kappa}_1+i \tilde{U}_1}{2}t)
       +
        R_1
        \sinh (\frac{\tilde{\kappa}_1+i \tilde{U}_1}{2}t)
        \right)^2}.
	\end{align}
    where $R_1 = (\kappa + i U(2\nth+1))/(\tilde{\kappa}_1+i \tilde{U}_1)$ in agreement with Ref.~\cite{Mark_Old_SPR_1984}. Here $\hat{r}_{m, \mu}$ and $\hat{l}_{m, \mu}$ are the right and left eigenvectors of $\hL$ with eigenvalue $\lambda_{m, \mu}$ (see SM \cite{SM}). Fourier-transforming Eq.~(\ref{eq:GR}) gives us the frequency-resolved Green's function $G^R[\omega]$, which can easily be accesssed in several experimental platforms.
    %The position and widths of the peaks in Fourier space tell us about the eigenvalues of $\hL$, whereas their height gives us information about the eigenvectors \cite{Scarlatella_NJP_2019}.
    In a similar manner, higher-order response functions can be directly tied to eigenvalues and eigenvectors for higher $m$ modes.  
    
    %    -in-$\ha$ and $\ha^\dagger$ correlation function probe higher $m$ modes.  %One applies a weak coherent tone with a frequency $\omega$ to the mode $\ha$ and measures the output field, which readily gives the response function $G^R[\omega]$ \cite{RMP_Clerk} .  The formal solution to the problem can thus be used as a precise way to measure the bare decay rate $\kappa$ and non-linearity $U$. We note that the physics of a damped quantum harmonic oscillator has recently received renewed interest in the literature \cite{Steele_Kerr_2021}.\alexcom{Awkward place to mention the Steele paper, not sure where else to do so.}\alexcom{Discuss various limits of $G^R(t)$? Probably not enough space. Would be also nice to write spectral decomposition, but again not enough space.}

    %%%%%%%%%%%
    %%%%%%%%%%%%%%%%%%%%%%%%%%%%%%%%%%%%%%%%%%%%%%%%%%%%%%
	%%%%%%%%%%%%%%%%%%%%%%%%%%%%%%%%%%%%%%%%%%%%%%%%%%%%%%
	%%%%%%%%%%%%%%%%%%%%%%%%%%%%%%%%%%%%%%%%%%%%%%%%%%%%%%
	
	While for clarity we have focused here on a single-mode problem, a completely analogous approach allows one to analytically diagonalize a truly many-body model, where we now have a set of bosonic modes, each with Kerr nonlinearities and thermal dissipation, coupled to one another via cross-Kerr interaction of the form $U_{ab} \ha^\dagger \ha \hat{b}^\dagger \hat{b}$.  As we show in the SM \cite{SM}, our method applies directly here: in each symmetry-constrained block, the non-trivial interaction terms become effectively quadratic.   We also show this setup remains solvable if we were to add dephasing to each mode (as described by the dissipators $2 \kappa_{\phi,j} \mathcal{D}[\ha_j^\dagger \ha_j] \hrho$).

    \textit{Dissipative Ising Model}.--- 
    We next show that our symmetry-based approach can be used for a completely different kind of system, namely a dissipative Ising model of $N$ spins. The Lindblad master equation reads
    \begin{align}\nonumber
        \partial_t \hrho
        &
        =
        -i[
        \sum_{j<k}
        J_{jk} \hat{\sigma}^z_j \hat{\sigma}^z_k
        +
        \sum_{j}
        h_j \hat{\sigma}^z_j
        ,
        \hrho
        ]
        +
        \sum_{j}
        \gamma_{-,j}
        \mathcal{D}[\hat{\sigma}^-_j] \hrho 
        \\
        &
        +
        \sum_{j}
        \gamma_{+,j}
        \mathcal{D}[\hat{\sigma}^+_j] \hrho 
        +
        \sum_{j}
        \gamma_{\phi,j}
        \mathcal{D}[\hat{\sigma}^z_j] \hrho 
        \equiv \mathcal{L}\hrho.
        \label{eq:SpinMasterEq}
    \end{align}
%    \alexcom{Label $\mathcal{L}$ differently to differentiate between oscilaltor and spin Lindbladian?}
It describes $N$ interacting two-level systems with arbitrary Ising couplings $J_{jk}$, each with its own local magnetic field $h_j$. Each spin is also subject to local spin relaxation, pumping, and dephasing characterized by the rates $\gamma_{-,j}$, $\gamma_{+,j}$ and $\gamma_{\phi,j}$ respectively.
Note that Ref.~\onlinecite{Ana_Maria_Diss_Ising_PRA_2013} was able to exactly calculate equal-time averages of one and two spin operators for this model, by using a stochastic unravelling of $\mathcal{L}$ and analytically performing the average over trajectories.  Our alternate approach goes much further: not only does it permit a simpler method for calculating averages, it also provides a direct means for obtaining the full dissipation spectrum, multi-time correlation functions and the full many-body density matrix.
% This same model was studied in Ref.~\onlinecite{Ana_Maria_Diss_Ising_PRA_2013}, where the authors were in principle able to compute arbitrary correlation functions using the quantum trajectory approach. Our solution not only provides a simpler and more intuitive way to compute these correlation functions, but also provides direct access to the full dissipation spectrum and the dynamics of the full many-body density matrix.
% ; this would be nearly impossible to obtain by averaging over stochastic trajectories. 
	
    $\mathcal{L}$ is invariant under arbitrary, independent local rotations around the $z$ axis of each spin, i.e.~$\hat{\sigma}_j^{\pm} \to e^{\pm i \theta_j} \hat{\sigma}_j^{\pm}$. There are thus $N$ weak $U(1)$ symmetries, one for each spin, generated by the superoperators $[\hat{\sigma}^z_j, \cdot]/2$. Each of these generators has two non-degenerate eigenvalues $m_j = \pm1$ whose eigenvectors are coherences $\ket{\uparrow_j}\bra{\downarrow_j} = \hat{\sigma}^{+}_j$ and $\ket{\downarrow_j}\bra{\uparrow_j} = \hat{\sigma}^{-}_j$.  There is also a two-fold degenerate eigenvalue $m_j = 0$ with associated population eigenvectors $\ket{\uparrow_j}\bra{\uparrow_j}$ and $\ket{\downarrow_j}\bra{\downarrow_j}$. The Lindbladian necessarily commutes with each generator and thus takes on a block diagonal form, where each block is indexed by $\vec{M} = \{m_1, \dots, m_N\}$, i.e.~the vector formed by the eigenvalues of the generators.  Given that the $m_j = \pm 1$ eigenvalues are non-degenerate whereas the $m_j = 0$ eigenvalues are two-fold degenerate, for a specific block indexed by $\vec{M}$, we can parition our spins into a a set of ``frozen" spins (i.e.~spins $j$ with $m_j = \pm 1$) and ``active" spins (i.e.~spins $j$ with $m_j = 0$).  Within the specific block described by a given $\vec{M}$, the populations of the active spins can fluctuate.  Formally, if we let $\hrho_{\vec{M}}$ denote the density matrix projected onto the subspace indexed by $\vec{M}$, then we have
    \begin{align} \nonumber
        \hrho_{\vec{M}}
        &
        =
        \hrho_{\rm froz}
        \times
        \hrho_{\rm act}
        \\ \label{eq:rho_M}
        &
        =
       \left(
        \prod_{j \: \rm{frozen}}
        \hat{\sigma}^{m_j}_j
        \right)
        \left(
        \sum_{\vec{s}_{\rm act}}
        P(
        \vec{s}_{\rm act}
        )
        \ket{\vec{s}_{\rm act}}\bra{\vec{s}_{\rm act}}
        \right)
    \end{align}
    where $\vec{s}_{\rm act} = \{s_j \: | \: j \: \rm{active} \}$ and $s_j \in \{\uparrow_j, \downarrow_j\}$. In each block $\hrho_{\vec{M}}$ factorizes as a product over coherences $\hrho_{\rm froz}$ and a \textit{classical} density matrix $\hrho_{\rm act}$ described entirely by a probability distribution $P(\vec{s}_{\rm act})$ for a ensemble of two-level systems. If we let $z(\vec{M})$ denote the number of zero eigenvalues of $\vec{M}$, which is by definition the number of active spins, then the size of the Lindblad block indexed by $\vec{M}$ is $2^{z(\vec{M})}$. 
    % \alexcom{Slight abuse of notation calling $P(\vec{s}_{\rm act})$ a probability distribution, but I think its fine.} 

    %For a given block indexed by $\vec{M}$, there are set of ``frozen" and ``active" spins. The frozen spins are those for which the eigenvalues are non-degenerate $m_j = \pm 1$, that is to say coherenecs, whereas the active ones are populations and correspond to the degenerate $m_j = 0$ eigenvalue. 
    
    Just as in the dissipative non-linear oscillator model, the existence of weak symmetry is not enough to make the system analytically solvable, as it only guarantees the block diagonal structure of Eq.~(\ref{eq:rho_M}). There are still many blocks whose dimension is exponentially large in the number of spins, encoding what would seem to be a complicated dissipative many-body problem. 
    Instead, further simplification emerges from the form of the interaction and the fact that a mean-field decoupling becomes exact in each symmetry sector.  
    %  Rather, the simplification emerges as a result of replacing the generator by its eigenvalue in each block. 
    We show in the SM \cite{SM} that, upon projecting into the subspace indexed by $\vec{M}$, this amounts to making the replacement
    \begin{align}\label{eq:Spin_MFT}
        [
        \sum_{j<k}
        J_{jk} \hat{\sigma}^z_j \hat{\sigma}^z_k
        ,
        \hrho
        ]
        \to
        \sum_{j}
        \{
        J_{j}^{\rm eff}(\vec{M}) \hat{\sigma}^z_j
        ,
        \hrho_{\vec{M}}
        \}
    \end{align}
    where we have defined $ J_j^{\rm eff}(\vec{M})  = \sum_{k \neq j} J_{jk} m_k$. Using Eq.~(\ref{eq:Spin_MFT}), we therefore see that within each block, mean-field theory becomes exact:
    the spin-spin interaction has been replaced by a (sector-dependent) static $z$ magnetic field on each spin,  $J_{j}^{\rm eff}(\vec{M})$. Combined with the local nature of the dissipation, it follows that the classical probability describing the active spin factorizes $\hrho_{\rm active} = \prod_{j \: \rm active}\left(p_{\uparrow, j } \ket{\uparrow_j}\bra{\uparrow_j} + p_{\downarrow, j} \ket{\downarrow_j}\bra{\downarrow_j} \right)$ and the equations of motion for the coefficients read
    \begin{align}\label{eq:EOM_Spin_Probabilities}
        \partial_t
        \begin{pmatrix}
            p_{\uparrow, j }
            \\
            p_{\downarrow, j }
        \end{pmatrix}
        =
        \begin{pmatrix}
        -2 i J^{\rm eff}_j - \gamma_{-,j}    & \gamma_{+,j} \\
        \gamma_{-, j} & 2 i J^{\rm eff}_j - \gamma_{+,j}
        \end{pmatrix}
        \begin{pmatrix}
        p_{\uparrow, j }
        \\
        p_{\downarrow, j }
        \end{pmatrix}
    \end{align}
      where, for the sake of compactness, we have dropped the $\vec{M}$ dependence of $J_{j}^{\rm eff}$.
    
    The above exact decoupling has thus allowed us to map a many-body quantum problem onto an effective classical model of non-interacting spins.  To see this explicitly, note that Eq.~(\ref{eq:EOM_Spin_Probabilities}) would correspond precisely to a classical master equation for a two-state system if not for the strange imaginary terms $\propto J_{j}^{\rm eff}$ on the diagonals. These terms also admit a simple classical interpretation. Consider the random variable $\hat{s}_j = \int_0^t dt' \hat{\sigma}^z_j(t') $, i.e.~the integral of the classical telegraph fluctuations of spin $j$.  
    We can now interpret $2J_j^{\rm eff}$ as a conjugate variable to this stochastic quantity (i.e.~a so-called ``counting field").   
    Viewed as a function of $2J_j^{\rm eff}$, the solution to Eq.~(\ref{eq:EOM_Spin_Probabilities}) allows us to obtain the time-dependent moment-generating function of $\hat{s}_j$, i.e. $\Lambda[2J_j^{\rm eff}] = \int d s_j P(s_j) e^{-2i J ^{\rm eff}_j s_j}$.  In a concrete sense, one concludes that the frozen spins are measuring the classical fluctuations of the active spins at a rate determined by $J_{ij}$.  The upshot is that our solution method can be viewed as having made a quantum-to-classical mapping in each symmetry-constrained block.

  The above exact decoupling of spins in each symmetry block immediately implies that all Liouvillian eigenvalues can be written as a sum over single-spin eigenvalues $\lambda_j(\vec{M})$.  A simple calculation yields
    \begin{align}\label{eq:Spin_Energies}
    \lambda_j(\vec{M})
    =
    \begin{cases}
    \mp i 2 h_j -\Gamma_j-2 \gamma_{\phi, j}, & j \: \rm{frozen}
    \\
    \Gamma_j
    \pm
    \sqrt{\Gamma_j^2-4 J^{\rm eff}_j(J^{\rm eff}_j+i \eta_j)},
    &
    j \: \rm{active}
    \end{cases}
\end{align}
with $\Gamma_j = (\gamma_{+,j}+\gamma_{-,j})/2$ and $\eta_j = (\gamma_{+,j} - \gamma_{-,j})/2$. Equation (\ref{eq:Spin_Energies}) tells us that coherences $\ket{\downarrow_j}\bra{\uparrow_j}$ and $\ket{\uparrow_j}\bra{\downarrow_j}$ behave as expected: they oscillate with a frequency controlled by the local magnetic field and decay at a rate set by the local dephasing and relaxation processes, independently of all other spins.  Populations however both decay and oscillate depending on the strength of the counting field $2J^{\rm eff}$ relative to the strength of the relaxation processes. The right and left eigenvectors factorize in a similar way, and one only needs to solve a $2 \times 2$ matrix eigenvalue problem to determine their form. As such, we leave those details to the SM \cite{SM}. 

With both the eigenvectors and eigenvalues in hand, we can again compute any physical quantity of interest for this model. In the SM \cite{SM}, we provide an example of this, for the case where all spins are initially all pointing along the $x$ direction.  As mentioned earlier, analogous quantities were calculated in  Ref.~\cite{Ana_Maria_Diss_Ising_PRA_2013} using an alternative method.
%based on a stochastic unravelling of the master equation.  Our solution however provides a much simpler route, as there is no stochastic averaging required.  Our method also provides the entire dissipation spectrum (something not easily done using the method of  Ref.~\cite{Ana_Maria_Diss_Ising_PRA_2013}), allowing one to identify new generic properties of the model.  
Our approach greatly simplifies the calculation, and also allows insights not possible using the trajectory method of Ref.~\cite{Ana_Maria_Diss_Ising_PRA_2013}, as we have access to the full dissipation spectrum.
For example, we find that our many-body Liouvillian can exhibit an exceptional point (EP) structure (see SM \cite{SM}), wherein the dynamics are exceptionally sensitive to small parameter changes. Such Lindblad EPs have been the subject of considerable recent interest  \cite{Kosloff_EP_PRE_2018, Nori_EP_PRA_2019, Nori_EP_PRA_2020}, though there are few truly many-body examples. Our approach can also be used to analytically find the full time-evolved many-body density matrix $\hrho(t)$ for an arbitrary initial condition (which would be difficult if not impossible to do using trajectories).  

Similar to our discussion of the dissipative nonlinear bosonic model earlier, we have for clarity sketched the simplest non-trivial dissipative spin model where our symmetry-based solution method holds.  The effective quantum-to-classical mapping we have established is in fact valid for a large class of dissipative spin models. For example, there are still $N$ weak $U(1)$ symmetries if we add to our model correlated spin-loss or flips for an arbitrarily large number of spins such as, e.g. $\mathcal{D}[\hat{\sigma}_j^- \hat{\sigma}_k^+]$. The block-diagonal decomposition Eq.~(\ref{eq:rho_M}) thus follows, as does the mean-field replacement Eq.~(\ref{eq:Spin_MFT}). The only difference is that classical probability distribution describing the active spins does not factorize; nevertheless the equations of motion in each block is exactly equivalent to a classical master equation of correlated spins with a counting field for each spin $J^{\rm eff}_j$.  This suggests that our approach could be a powerful means to attack a range of dissipative spin models.   

% \alexcom{Need a good way to end this paragraph, not sure what more to say though.} \alexcom{Make connection to quantum-to-classical mapping with Kerr here or earlier?}

 %%%%%%%%%%%%%%%%%%%%%%%%%%%%%%%%%%%%%%%%%%%%%%%%%%%%%%
	%%%%%%%%%%%%%%%%%%%%%%%%%%%%%%%%%%%%%%%%%%%%%%%%%%%%%%
	%%%%%%%%%%%%%%%%%%%%%%%%%%%%%%%%%%%%%%%%%%%%%%%%%%%%%%
    
    \textit{Conclusion}.--- 
    Our work shows how continuous weak symmetries can enable the analytic solution of a wide class of interacting dissipative quantum models.  While we analyzed to specific examples (one bosonic, the other spin-based), we stress that the method could be applied to a variety of other systems.  It also provides a powerful starting point for systematic approximation methods for systems with additional terms that break the relevant weak symmetry.  For example, as our approach provides simple analytic expressions for all eigenvalues and eigenvectors, it could be directly combined with Lindblad perturbation theory \cite{Li2014, LiPRX2016, Ryo_2021_Keldysh_CQED}.  In future work, it would be interesting to reformulate the general structure we have exploited here in terms of a dissipative Keldysh action \cite{Diehl_Keldysh_Review_2016, Kamenev_Book_2011}; this could enable an extension of our method to non-Markovian dissipative systems.
    
    This work is supported by the Air Force Office of Scientific Research MURI program under Grant No. FA9550-19-1-0399, and by the Simons Foundation (Award No.~669487, AC).

\bibliography{Weak_Symmetries.bib}
 \end{document}

\title{Supplementary Materials: Exact solutions of interacting dissipative systems via weak symmetries}

	\author{Alexander McDonald}
	\affiliation{Pritzker School of Molecular Engineering, University of Chicago, Chicago, IL 60637, USA}
	\affiliation{ Department of Physics, University of Chicago, Chicago, IL 60637, USA}
	\author{Aashish A. Clerk}
	\affiliation{Pritzker School of Molecular Engineering, University of Chicago, Chicago, IL 60637, USA}

	\date{\today}

\maketitle

\setcounter{equation}{0}
\setcounter{figure}{0}
\setcounter{table}{0}
\setcounter{page}{1}
\setcounter{section}{0}
\renewcommand{\theequation}{S\arabic{equation}}
\renewcommand{\thefigure}{S\arabic{figure}}
\renewcommand{\thetable}{S\arabic{table}}
\renewcommand{\bibnumfmt}[1]{[S#1]}
\renewcommand{\citenumfont}[1]{S#1}

\maketitle

%%%%%%%%%%%%%%%%%%%%%%%%%%%%%%%%%%%%%
%%%%%%%%%%%%%%%%%%%%%%%%%%%%%%%%%%%%%
\section{Eigenvectors and eigenvalues of incoherently-driven non-linear oscillator Lindbladian}\label{SM:Eigenvales_eigenvectors}
In this section, we find the eigenvectors and eigenvalues of the incoherently-driven non-linear oscillator, whose third quantized form is given by Eqs~(1)-(2) in the main text. As we then explain in the subsequent discussion following this equation, one can make use of the $U(1)$ symmetry to reduce the problem to a quadratic one in each block $\hL_m$ by making the substitution $\haL^\dagger \haL-\haR \haR^\dagger \to m$. Using commutation relations $[\haL, \haL^\dagger] = -[\haR, \haR^\dagger] = \hone$ along with the fact that any superoperator which acts on the left commutes with those acting on the right, we get
\begin{align}
    \hL_m
    =
    \left(-i(\omega_0-U)m+\frac{\kappa}{2}\right)\hone
    +
	\kappa(\nth+1)\haR^\dagger \haL 
	+ 
	\kappa \nth \haL^\dagger \haR 
	+
    (-i\frac{U}{2}m-\frac{\kappa}{2}(2\nth+1))
	\left(
	\haL^\dagger \haL
	+
	\haR^\dagger \haR \right). 
\end{align}
Following standard third-quantization, we first want to find a set of superoperators which satisfy
\begin{align}\label{SM:Eigen_1}
    &[\hL_m, \hcp^\dagger] =  -\frac{\Gamma_m}{2} \hcp^\dagger,
    \hspace{0.5 cm}
    [\hL_m, \hdp] = \frac{\Gamma_m}{2}\hdp
    \\ \label{SM:Eigen_2}
    &
    [\hL_m, \hcm] =  -\frac{\Gamma_m}{2} \hcm,
    \hspace{0.5 cm}
    [\hL_m, \hdm^\dagger] = \frac{\Gamma_m}{2} \hdm^\dagger.
\end{align}
In addition, the only non-vanishing commutation relation between these superoperators should be
\begin{align}\label{SM:Superoperator_Commutation}
    [
    \hdp,
    \hcp^\dagger
    ]
    =
    [
    \hcm,
    \hdm^\dagger
    ]
    =
    \hone.
\end{align}
Making an ansatz that $\hcm$ and $\hdp$ are linear combinations of $\haL$ and $\haR$ whereas $\hcp^\dagger$ and $\hdm^\dagger$ are linear combinations of $\haL^\dagger$ and $\haR^\dagger$, we can use the superoperator commutation relation to turn Eqs.~(\ref{SM:Eigen_1})-(\ref{SM:Eigen_2}) into a $2 \times 2$ matrix eigenvalue problem. These can be readily solved and gives
	\begin{align}
	    \Gamma_m 
	    =
	    \sqrt{
	    \kappa^2
	    -
	    U^2m^2
	    +
	    2 i \kappa U m (2\nth+1)
	    }
	\end{align}
Note that we always choose the branch cut of the square root such that $\text{Re} \: \Gamma_m > 0$. We can then define
\begin{align}\label{SM:Def_Eigen_One}
    &\begin{pmatrix}
     \hdp \\ \hcm 
    \end{pmatrix}
    =
    \frac{1}{\sqrt{2}}
    \begin{pmatrix}
    \frac{1}{2 \Gamma_m}(i U m+ (\Gamma_m+\kappa)+2 \kappa(2\nth +1))
    &
    \frac{-1}{2 \Gamma_m}(i U m-(\Gamma_m+\kappa)+2 \kappa(2\nth +1))
    \\
    \frac{1}{\Gamma_m+\kappa}
    (\Gamma_m+\kappa-i U m) 
    &
    \frac{-1}{\Gamma_m+\kappa}
    (\Gamma_m+\kappa+i U m)
    \end{pmatrix}
    \begin{pmatrix}
     \haL \\ \haR 
    \end{pmatrix}
    \\ \label{SM:Def_Eigen_Two}
    &
    \begin{pmatrix}
     \hdm^\dagger \\ \hcp^\dagger 
    \end{pmatrix}
    =
    \frac{1}{\sqrt{2}}
    \begin{pmatrix}
    \frac{-1}{2 \Gamma_m}(i U m - (\Gamma_m+\kappa)+2 \kappa(2\nth +1))
    &
    \frac{1}{2 \Gamma_m}(i U m+(\Gamma_m+\kappa)+2 \kappa(2\nth +1))
    \\
    \frac{1}{\Gamma_m+\kappa}
    (\Gamma_m+\kappa+i U m) 
    &
    \frac{-1}{\Gamma_m+\kappa}
    (\Gamma_m+\kappa-i U m)
    \end{pmatrix}
    \begin{pmatrix}
     \haL^\dagger \\ \haR^\dagger 
    \end{pmatrix}
\end{align}
as the superoperators which satisfy Eqs.~(\ref{SM:Eigen_1}-\ref{SM:Superoperator_Commutation}). Expressing $\hL_m$ in this new basis, we obtain
\begin{align}\label{SM:Compact_Lm}
    \hL_m
    =
    \left((\omega_0-U)m-\frac{i(\Gamma_m-\kappa)}{2}\right)\hone
    -
    \frac{i\Gamma_m}{2}
    \left(
    \hcp^\dagger \hdp-\hcm \hdm^\dagger
    \right).
\end{align}
We now want to find the right and left ``vacuum'' of each sector, that is to say operators which satisfy
\begin{align}\label{SM:Right_Vacuum}
    &\hdp \ket{\hat{0}^r_m}
    =
    \hdm^\dagger \ket{\hat{0}^r_m}
    =
    0,
    \\ \label{SM:Left_Vacuum}
    &
    \bra{\hat{0}^l_m} \hcp^\dagger
    =
    \bra{\hat{0}^l_m} \hcm
    =
    0.
\end{align}
Using the definitions Eqs.~(\ref{SM:Def_Eigen_One})-(\ref{SM:Def_Eigen_Two}), one can easily demonstrate show that these states are Gaussian
\begin{align}
    \hat{0}_m^r
    &=
    (1-e^{-\beta_m^r} (e^{-\beta_m^l})^*)
    \sum_{n=0}^\infty
    e^{-\beta^r_m n}
    \ket{n}\bra{n}
    \\
    (\hat{0}_m^l)^\dagger
    &=
    \sum_{n=0}^\infty
    (e^{-\beta^l_m n})^*
    \ket{n}\bra{n}
\end{align}
with complex Boltzmann factors defined as
\begin{align}\label{SM:Boltz_R}
    e^{-\beta_m^r}
    &=
    \frac{
    i U m-(\Gamma_m+\kappa)+2 \kappa(2\nth+1)
    }
    {
    i U m+(\Gamma_m+\kappa)+2 \kappa(2\nth+1)
    }
    ,
    \\ \label{SM:Boltz_L}
    (e^{-\beta_m^l})^*
    &=
    \frac
    {
    \Gamma_m+\kappa- i U m
    }
    {
    \Gamma_m+\kappa+i U m
    }.
\end{align}
The normalization is chosen such that $ \langle \hat{0}^l_m | \hat{0}^r_m \rangle = \Tr  (\hat{0}^l_m)^\dagger \hat{0}^r_m  = 1$. Further, note that in the limit $U \to 0$ (or in the $m = 0$ sector) we recover $\hat{0}^r_m \to \hrho_{\rm ss}, \hat{0}^l_m \to \hat{1}$ where $\hrho_{\rm ss}$ is the steady state, i.e. the Gaussian state with an average occupation $\nth$.

With these ``vacuum'' states and the compact form of $\hL_m$ provided in Eq.~(\ref{SM:Compact_Lm}), we are now in a position to describe the eigenvectors and eigenvalues of $\hL$.
They are given by
	\begin{align}\label{SM:Right_Eigen_Def}
	    \ket{\hat{r}_{m,\mu}}
	    &=
	    \frac{1}{\sqrt{\mu!(\mu+|m|)!}}
	    \begin{cases}
	    (\hcp^\dagger \hcm)^\mu(\hcp^\dagger)^m
	    \ket{\hat{0}_m^r}
	    ,
	    &
	    m \geq 0
	    \\
	    (\hcp^\dagger \hcm)^\mu(-\hcm)^{-m}
	    \ket{\hat{0}_m^r}
	    ,
	    &
	    m < 0
	    \end{cases}
	    \\ \label{SM:Left_Eigen_Def}
	    \langle \hat{l}_{m,\mu} |
	    &=
	    \frac{1}{\sqrt{\mu!(\mu+|m|)!}}
	    \begin{cases}
	    \bra{\hat{0}_m^l}
	    (\hdp)^m(-\hdm^\dagger \hdp)^\mu
	    ,
	    &
	    m \geq 0
	    \\
	    \bra{\hat{0}_m^l}
	    (\hdm^\dagger)^{-m}(-\hdm^\dagger \hdp)^\mu
	    ,
	    &
	    m < 0
	    \end{cases}
	\end{align}
and have the corresponding eigenvalues are
\begin{align}\label{SM:Oscillator_Eigenvalues}
    \lambda_{m,\mu}
    =
    -i(\omega_0-U) m
    -
    \frac{(\Gamma_m-\kappa)}{2}
    -\frac{\Gamma_m}{2}
    \left(
    |m|+2\mu
    \right).
\end{align}
which is identical to the result in the main text once we correctly identify the renormalized decay rate and non-linearity as $\tilde{\kappa}_m = \Re \Gamma_m$ and $\tilde{U}_m = \Im \Gamma_m$. To arrive at this result, we must first remember that by definition $\hL_m$ is the full Lindbladian projected onto the subspace spanned by eigenvectors of $\haL^\dagger \haL-\haR \haR^\dagger$ with an eigenvalue of $m$. Using the commutation relations
\begin{align}\label{SM:Lower}
    &[\haL^\dagger \haL-\haR \haR^\dagger
    , \boldsymbol{\hat{a}}_{\rm L/R}]
    =
    -\boldsymbol{\hat{a}}_{\rm L/R}
    \\\label{SM:Raise}
    &
    [\haL^\dagger \haL-\haR \haR^\dagger
    , \boldsymbol{\hat{a}}_{\rm L/R}^\dagger]
    =
    \boldsymbol{\hat{a}}_{\rm L/R}^\dagger
\end{align}
we see that $\boldsymbol{\hat{a}}_{\rm L/R}$ and $\boldsymbol{\hat{a}}_{\rm L/R}^\dagger$ lowers or raises the eigenvalue $m$ by one respectively. The same is then true of $\hcm, \hdp$ and $\hcp^\dagger, \hdm^\dagger$ respectively. Both $\ket{\hat{0}^r_m}$ and $\langle \hat{0}^l_m|$ are a linear combination of Fock state projectors, and they are therefore elements of the $m = 0$ subspace. By construction, the left and right eigenvectors thus satisfy $(\haL^\dagger \haL - \haR \haR^\dagger)\ket{\hat{r}_{m,\mu}} = m \ket{\hat{r}_{m,\mu}} $ and  $\langle \hat{l}_{m,\mu} | (\haL^\dagger \haL - \haR \haR^\dagger) = m \langle \hat{l}_{m,\mu} |$. It then follows that
\begin{align}
    \hL  \ket{\hat{r}_{m,\mu}} 
    &=
    \hL_m \ket{\hat{r}_{m,\mu}} 
    \\
    \langle \hat{l}_{m,\mu} | \hL   
    &=
    \langle  \hat{l}_{m,\mu} | \hL_m.
\end{align}
We can then use Eq.~(\ref{SM:Compact_Lm}), the commutation relations Eq.~(\ref{SM:Superoperator_Commutation}) and the defining property of $\ket{\hat{0}^r_m}$ and $\langle \hat{0}^l_m|$ in Eqs.~(\ref{SM:Right_Vacuum})-(\ref{SM:Left_Vacuum}) to show that $\ket{\hat{r}_{m,\mu}}$ and $\langle \hat{l}_{m,\mu} |$ are the eigenvectors of $\hL$ with eigenvalues $\lambda_{m,\mu}$. Further, using the same commutation relations in Eq.~(\ref{SM:Superoperator_Commutation}) and the normalization $\langle \hat{0}^l_m | \hat{0}^r_m \rangle = 1$ one can show the normalization convention we have chosen is such that the biorthogonality condition
\begin{align}
    \langle \hat{l}_{m',\mu'} | \hat{r}_{m,\mu} \rangle
    =
    \delta_{m ,m'}
    \delta_{\mu,\mu'}
\end{align}
is satisfied. 

\begin{figure}[t]
	\centering
    \includegraphics[width=0.45 \textwidth]{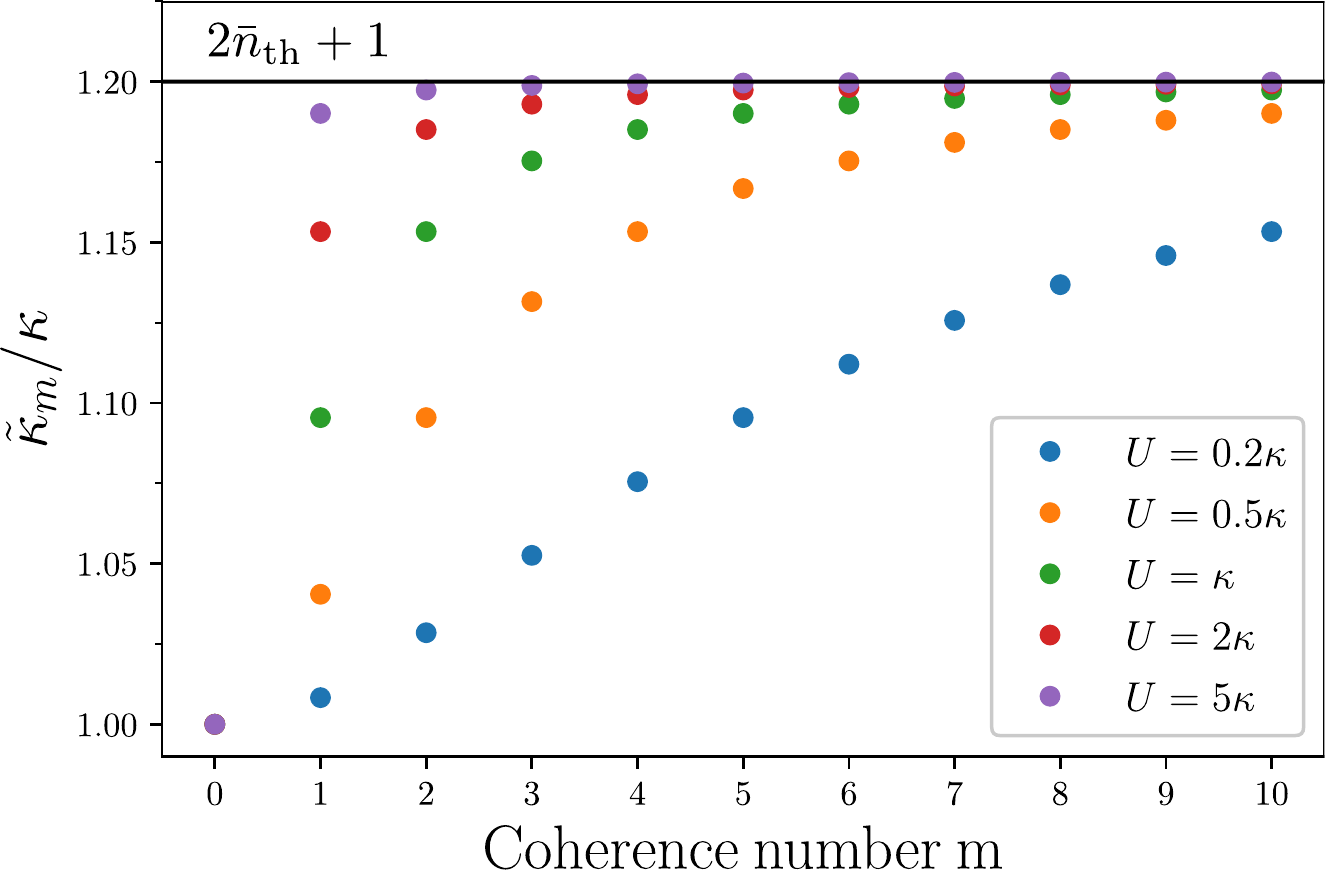}
	\caption{Renormalized decay rate $\tilde{\kappa}_m$ as a function of $m$ for various values of $U$ and $\nth = 0.1$. For $\kappa \ll U m$, one reaches the upper bound of the renormalized decay rate $ \tilde{\kappa}_m \lesssim \kappa(2\nth+1)$ (see Sec.~I of the SM for details).
	}
	\label{fig:Kappa}
	\end{figure}

Note that we readily obtain a simple intuituve result rule by Taylor expanding the eigenvalues and eigenvectors to first order in $\kappa$ and $\nth$. We first note that when $\nth \to 0$ and $\nth$, the ``vacuum" states in each sector coincide with the vacuum state projector
\begin{align}
    \lim_{\nth \to 0, \kappa \to 0}
    \hat{0}_m^r
    =
    \lim_{\nth \to 0, \kappa \to 0}
    (\hat{0}_m^l)^\dagger
    =
    \ket{0}\bra{0}.
\end{align}
Another simple calculation yields, in the same limit,
\begin{align}\label{SM:Def_Eigen_Limit}
    &
    \lim_{\nth \to 0, \kappa \to 0}
    \begin{pmatrix}
     \hdp \\ \hcm 
    \end{pmatrix}
    =
    \begin{pmatrix}
    \frac{1}{\sqrt{2}} & 0 \\
    0 & -\sqrt{2} 
    \end{pmatrix}
    \begin{pmatrix}
     \haL \\ \haR 
    \end{pmatrix}
    \\ \label{SM:Def_Eigen_Limit_2}
    &
    \lim_{\nth \to 0, \kappa \to 0}
    \begin{pmatrix}
     \hdm^\dagger \\ \hcp^\dagger 
    \end{pmatrix}
    =
    \begin{pmatrix}
    0 & \frac{1}{\sqrt{2}} \\
    \sqrt{2} & 0
    \end{pmatrix}
    \begin{pmatrix}
     \haL^\dagger \\ \haR^\dagger 
    \end{pmatrix}.
\end{align}
Using the definition of the right and left eigenvectors Eqs.~(\ref{SM:Right_Eigen_Def})-(\ref{SM:Left_Eigen_Def}), we recover the fact that the right and left eigenvectors are Fock state projectors (up to an irrelevant normalization factor):
	\begin{align}\label{SM:Right_Eigen_Limit}
	    \lim_{\nth \to 0, \kappa \to 0} 
	    \hat{r}_{m,\mu}
	    &=
	    (-1)^\mu(\sqrt{2})^{2\mu+|m|}
	    \begin{cases}
	    \ket{\mu + |m|}\bra{\mu}
	    ,
	    &
	    m \geq 0
	    \\
	    \ket{\mu}\bra{\mu + |m|}
	    ,
	    &
	    m < 0
	    \end{cases}
	    \\ \label{SM:Left_Eigen_Limit}
	    \lim_{\nth \to 0, \kappa \to 0}
	    \hat{l}_{m,\mu}^\dagger
	    &=
	    (-1)^\mu(\sqrt{2})^{-2\mu-|m|}
	    \begin{cases}
	    \ket{\mu}\bra{\mu + |m|}
	    ,
	    &
	    m \geq 0
	    \\
	    \ket{\mu+|m|}\bra{\mu}
	    ,
	    &
	    m < 0
	    \end{cases}
	\end{align}
 We can then use Lindblad perturbation theory \cite{Li2014, LiPRX2016} to compute the decay rate $-\text {Re} \lambda_{m, \mu}$ to first order in $\kappa$. If we let $\boldsymbol{\mathcal{\hat{V}}}$ denote the part of the full Lindblad superoperator which describes dissipation, then can write the first order correction to the eigenvalues in complete analogy to standard first-order perturbation theory
 \begin{align} \nonumber
     \text {Re} \lambda_{m, \mu}
     &
     =
     \langle \hat{l}_{m,\mu} |
     \boldsymbol{\mathcal{\hat{V}}}
      \ket{\hat{r}_{m,\mu}}
      +
      \mathcal{O}(\kappa^2)
      \\ \nonumber 
      &
      =
      -\frac{\kappa}{2}(\nth+1)
      \Tr
      \left[
      \ket{\mu}\bra{\mu + |m|}
      \{
      \ha^\dagger \ha, 
      \ket{\mu + |m|}\bra{\mu}
      \}
      \right]
    -\frac{\kappa}{2}\nth
      \Tr
      \left[
      \ket{\mu}\bra{\mu + |m|}
      \{
      \ha \ha^\dagger, 
      \ket{\mu + |m|}\bra{\mu}
      \}
      \right]
      +
      \mathcal{O}(\kappa^2)
      \\
      &
      =
      -\kappa (\nth+1)
      \frac{2\mu+|m|}{2}
      -
      \kappa \nth 
      (
      \frac{2\mu+|m|}{2}
      +
      1
      )
      +
      \mathcal{O}(\kappa^2)
 \end{align}
which agrees with Eq.~(\ref{SM:Oscillator_Eigenvalues}) upon expanding $\Gamma_m$ to first order. The first term is simply the average of the Fermi's Golden rule decay rate of the Fock states $\ket{\mu+|m|}$ and $\ket{\mu}$ with the appropriate stimulated emission factor. The second term is the average decay rate of $\ket{\mu+|m|+1}$ and $\ket{\mu+1}$ with the stimulated absorption term $\nth$. We also note, as we do in the main text, that this provides an upper bound to the renormalized sector-dependent decay rates $\tilde{\kappa}_m$ as show in Fig.~\ref{fig:Kappa}

    %%%%%%%%%%%
	% FIG SM 1 Kappa
	%%%%%%%%%%%
    %%%%%%%%%%%%%%%%%%%%%%%%%%%%%%%%%%%%%
    %%%%%%%%%%%%%%%%%%%%%%%%%%%%%%%%%%%%%
\section{Retarded response function}
Here we analytically find the retarded Green's function of the thermal Kerr oscillator. If we recall the definition of the third-quantized inner product $\langle \hat{B} | \hat{A} \rangle \equiv \Tr \hat{B}^\dagger \hat{A}$ and the annihilation and creation superoperators $\haL \ket{\hrho} \equiv \ket{\ha \hrho}$, $\haR \ket{\hrho} \equiv | \hrho \ha \rangle$, $\haL^\dagger \ket{\hrho} \equiv |\ha^\dagger \hrho \rangle$, and $\haR^\dagger \ket{\hrho} \equiv |\hrho \ha^\dagger\rangle$ then with the help of the quantum regression theorem we can write
\begin{align}
    G^R(t)
    &
    =
    -i \Theta(t)
    \langle
    [
    \ha(t)
    ,
    \ha^\dagger(0)
    ]
    \rangle
    \\
    &
    =
    -i \Theta(t)
    \left(
    \langle \hat{1} |  
    \haL e^{\hL t} \haL^\dagger 
    |\hrho_{\rm ss}\rangle
    -
    \langle \hat{1} |  
    \haR e^{\hL t} \haR^\dagger 
    |\hrho_{\rm ss}\rangle
    )
    \right)
\end{align}
Using the spectral decomposition of $\hL$, we can write the propegator as
\begin{align}
    e^{\hL t}
    =
    \sum_{m = -\infty}^{\infty}
    \sum_{\mu = 0}^\infty
    e^{\lambda_{m, \mu }t}
    | \hat{r}_{m,\mu} \rangle
    \langle \hat{l}_{m,\mu} |
\end{align}
Recall the both $|\hrho_{\rm ss} \rangle$ and $\langle \hat{1} |$, being a sum of Fock state projectors, are elements of the $m = 0$ subspace. From Eqs.~(\ref{SM:Raise})-(\ref{SM:Lower}), it follows that $\boldsymbol{\hat{a}}_{\rm L/R}^\dagger |\hrho_{\rm ss} \rangle$ and $\langle \hat{1} | \boldsymbol{\hat{a}}_{\rm L/R}]$ belong in the $m = 1$ subspace and are thus orthogonal to all right and left eigenvectors for which $m \neq 1$. Using Eqs.~(\ref{SM:Right_Eigen_Def})-(\ref{SM:Left_Eigen_Def}) we have
\begin{align}\nonumber
    G^R(t)
    &=
    -i \Theta(t)
    \sum_{\mu  = 0}^\infty
    e^{\lambda_{1, \mu }t}
    \left(
    \langle \hat{1} | \haL | \hat{r}_{1,\mu} \rangle
    \langle \hat{l}_{1,\mu} | \haL^\dagger |\hrho_{\rm ss}\rangle
    -
    (\haL \to \haR)
    \right)
    \\ \label{SM:Green's_Function_Intermediate_Step}
    &=
    -i \Theta(t)
    \sum_{\mu  = 0}^\infty
    \frac{e^{\lambda_{1, \mu }t} (-1)^\mu}{\mu!(\mu+1)!}
    \left(
    \langle \hat{1} | 
    \haL 
    (\boldsymbol{\hat{c}}_{+, 1}^\dagger)^{\mu+1}
    (\boldsymbol{\hat{c}}_{-, 1})^{\mu}
    |\hat{0}^{r}_{1} \rangle
    \langle \hat{0}^{l}_{1} |
    (\boldsymbol{\hat{d}}_{+, 1})^{\mu+1}
    (\boldsymbol{\hat{d}}_{-, 1}^\dagger)^{\mu}
    \haL^\dagger 
    |\hrho_{\rm ss}\rangle
    -
    (\haL \to \haR)
    \right).
\end{align}
To evaluate these matrix elements, we must use the fact that $\hat{0}^r_m$ and $\hat{0}^l_m$ are generalized Gaussian operators. Thus, one can use the Baker-Campbell-Hausdorff formula (or via direct computation) to write
\begin{align}
    \haR^\dagger 
    | \hat{0}^r_m \rangle
    =
    | \hat{0}^r_m \ha^\dagger \rangle
    =
    e^{-\beta^r_m}
    | \ha^\dagger \hat{0}^r_m  \rangle
    =
     e^{-\beta^r_m}
     \haL^\dagger
     | \hat{0}^r_m \rangle
     \\
     \langle \hat{0}^l_m | \haR
     =
     \langle \hat{0}^l_m \ha^\dagger |
     =
      \langle \ha \hat{0}^l_m| (e^{-\beta_m^l})^*
      =
      \langle  \hat{0}^l_m |
      \haL (e^{-\beta_m^l})^*
\end{align}
where the complex Boltzmann factors are defined in Eqs.~(\ref{SM:Boltz_R})-(\ref{SM:Boltz_L}). From the definition of $\hcm, \hcp^\dagger, \hdp$ and $\hdm^\dagger$ we then have
\begin{align}
     \langle \hat{1} | 
    \haL 
    (\boldsymbol{\hat{c}}_{+, 1}^\dagger)^{\mu+1}
    (\boldsymbol{\hat{c}}_{-, 1})^{\mu}
    |\hat{0}^{r}_{1} \rangle
    =
    \left(\frac{-i \sqrt{2}U}{\Gamma_1+\kappa}\right)^\mu
    \left(
    \frac{ \Gamma_1+\kappa+i U}{\sqrt{2}(\Gamma_1+\kappa)}
    \left[
   1
    -
    e^{-\beta^r_1}(e^{-\beta^l_1})^*
    \right]
    \right)^{\mu+1}
    \langle
    (\ha^\dagger)^{\mu+1}
    |
    (\ha^\dagger)^{\mu+1}
    \hat{0}^r_1
    \rangle
\end{align}
\begin{align}\nonumber
    \langle \hat{0}^{l}_{1} |
    (\boldsymbol{\hat{d}}_{+, 1})^{\mu+1}
    (\boldsymbol{\hat{d}}_{-, 1}^\dagger)^{\mu}
    \haL^\dagger 
    |\hrho_{\rm ss}\rangle
    &=
    \left(
    \frac{1}{\sqrt{2}2\Gamma_1}
    \left[
    i U+\Gamma_1+\kappa +2 \kappa(2\nth+1)
    \right]
    (
    1-e^{-\beta^r_1}(e^{-\beta^l_1})^*
    )
    \right)^{\mu+1}
    \\
    &
    \times
    \left(
    \frac{-1}{\sqrt{2}2\Gamma_1}
    \left[
    i U-\Gamma_1-\kappa +2 \kappa(2\nth+1)
    \right]
    (
    1-e^{\beta^r_1}e^{\beta}
    )
    \right)^{\mu}
    \langle 
    (\ha^\dagger)^{\mu+1}\hat{0}^l_1
    |
    (\ha^\dagger)^{\mu+1}\hrho_{\rm ss}
    \rangle
\end{align}
We can also readily compute the remaining overlaps as
\begin{align}
        \langle
    (\ha^\dagger)^{\mu+1}
    |
    (\ha^\dagger)^{\mu+1}
    \hat{0}^r_1
    \rangle
    =
    \Tr
    \left(
    \ha^{\mu+1}
    (\ha^\dagger)^{\mu+1}
    \hat{0}^r_1
    \right)
    =
    \sum_{n=0}^\infty
    \frac{(n+1)\cdots(n+\mu+1)
    e^{-\beta^r_1 n}}{(1-e^{-\beta^r_1}(e^{-\beta^l_1})^*)^{-1}}
    =
   (\mu+1)!
   \frac{  (1-e^{-\beta^r_1}(e^{-\beta^l_1})^*)}
   {(1-e^{-\beta^r_1})^{\mu+2}}
\end{align}
\begin{align}
    \langle 
    (\ha^\dagger)^{\mu+1}\hat{0}^l_1
    |
    (\ha^\dagger)^{\mu+1}\hrho_{\rm ss}
    \rangle
    =
    \Tr
    \left(
    (\hat{0}^l_1)^\dagger \ha^{\mu+1}
    (\ha^\dagger)^{\mu+1}
    \hrho_{\rm ss}
    \right)
    =
    (\mu+1)!
    \frac{1-e^{-\beta}}{(1-e^{-\beta}(e^{-\beta_1^l})^*)^{\mu+2}}
\end{align}
Putting all of this together after some straightforward but algebra, we obtain
\begin{align}
     \langle \hat{1} | 
    \haL 
    (\boldsymbol{\hat{c}}_{+, 1}^\dagger)^{\mu+1}
    (\boldsymbol{\hat{c}}_{-, 1})^{\mu}
    |\hat{0}^{r}_{1} \rangle
    \langle \hat{0}^{l}_{1} |
    (\boldsymbol{\hat{d}}_{+, 1})^{\mu+1}
    (\boldsymbol{\hat{d}}_{-, 1}^\dagger)^{\mu}
    \haL^\dagger 
    |\hrho_{\rm ss}\rangle
    =
   ((\mu+1)!)^2
   \frac{4 (\nth+1)}
   {(
   1
   +
   R_1
   )^2
   }
   \left(
   \frac{1-R_1}{1+R_1}
   \right)^\mu
\end{align}
where, as in the main text,
\begin{align}
    R_1
    =
    \frac{\kappa + i U(2\nth+1)}{\tilde{\kappa}_1+i \tilde{U}_1}
\end{align}
with $\kappa_1 = \text{Re} \: \Gamma_1$ and $\tilde{\Gamma}_1 = \text{Im} \: \Gamma_1$. Using $\langle \hat{1} | \haL = \langle \hat{1} | \haR $ and $\haR^\dagger| \hrho_{\rm ss} \rangle = \nth/(\nth+1) \haL^\dagger |\hrho_{\rm ss}\rangle $ we also immediately obtain
\begin{align}
        \langle \hat{0}^{l}_{1} |
    (\boldsymbol{\hat{d}}_{+, 1})^{\mu+1}
    (\boldsymbol{\hat{d}}_{-, 1}^\dagger)^{\mu}
    \haL^\dagger 
    |\hrho_{\rm ss}\rangle
    =
    ((\mu+1)!)^2
   \frac{4 \nth}
   {(
   1
   +
   R_1
   )^2
   }
   \left(
   \frac{1-R_1}{1+R_1}
   \right)^\mu.
\end{align}
Thus, the retarted Green's function takes the form
\begin{align}\nonumber
    G^R(t)
    &=
    -i \Theta(t)
    e^{-i(\omega_0-U) t+ \frac{\kappa}{2}t}
    \frac{4}{(R_1+1)^2}
    \sum_{\mu = 0}^\infty
    (\mu+1)e^{-(\tilde{\kappa}_1+i \tilde{U}_1)(\mu+1) t}
    \left(
    \frac{R_1-1}{R_1+1}
    \right)^\mu
    \\
    &
    =
    -i \Theta(t)
    \frac{e^{-i(\omega_0-U)t + \frac{\kappa}{2}t}}
    {
    \left(
    \cosh (\frac{\tilde{\kappa}_1+i \tilde{U}_1}{2}t)
    +
    R_1
    \sinh (\frac{\tilde{\kappa}_1+i \tilde{U}_1}{2}t)
    \right)^2}
\end{align}
which is identical with Eq.~(11) in the main text, and matches the result in Ref.~\cite{Mark_Old_SPR_1984}.
%%%%%%%%%%%%%%%%%%%%%%%%%%%%%%%%%%%%%
%%%%%%%%%%%%%%%%%%%%%%%%%%%%%%%%%%%%%
\section{Extending the method to interacting oscillators and dephasing}
Here, we will briefly explain how our solution technique is also applicable to the case where several oscillators are subject to a density-density interaction. For the sake of clarity, we will focus on the case where there are only two oscillators with mode annihilation operators $\ha$ and $\hb$; the generalization to an arbitrary number of oscillators, readily follows. The master equation reads
\begin{align}
    \partial_t \hrho
    =
    (
    \mathcal{L}_a
    +
    \mathcal{L}_b)
    \hrho
    -i
    U_{ab}
    [
     \ha^\dagger \ha \hb^\dagger \hb
    ,
    \hrho
    ]
\end{align}
where $\mathcal{L}_a$ and $\mathcal{L}_b$ are the superoperators which describe the uncoupled non-linear oscillators as in Eqs~(1)-(2) of the main text, which we already know how to diagonalize. Crucially the weak symmetries that allowed us to solve the $U_{ab} = 0$ case are still present. The eigenstates are thus labelled by $m_a$ and $m_b$, which are the integer eigenvalues of the superoperators  $[\ha^\dagger \ha, \cdot]$ and $[\hb^\dagger \hb, \cdot]$ respectively. To see why this proves useful, note that we can use standard commutator rules to write the density-density interaction as
\begin{align}\label{SM:Factorization}
    U_{ab}
    [
     \ha^\dagger \ha \hb^\dagger \hb
    ,
    \hrho
    ]
    =
    \frac{U_{ab}}{2}
    \left(
    \{ 
    \ha^\dagger \ha
    ,
    [\hb^\dagger \hb, \hrho]
    \} 
    +
    \{ 
    \hb^\dagger \hb
    ,
    [\ha^\dagger \ha, \hrho]
    \} 
    \right)
\end{align}
Projecting Eq.~(\ref{SM:Factorization}) onto the fixed $m_a$, $m_b$ subspace $\hrho_{m_a, m_b}$ we then obtain
\begin{align}
      U_{ab}
    [
     \ha^\dagger \ha \hb^\dagger \hb
    ,
    \hrho
    ]
    \to
    \frac{U_{ab} m_b}{2}
    \{\ha^\dagger \ha, \hrho_{m_a,m_b}\}
    +
    \frac{U_{ab} m_a}{2}
    \{\hb^\dagger \hb, \hrho_{m_a,m_b} \}
\end{align}
 There are thus two salient features that emerge once we coupled several oscillators via a density-density interaction. The first is that once we project onto the blocks indexed by the coherence numbers, the problem reduces to an effectively quadratic one. Further, we also see that mean field theory becomes exact within each block: the $a$ oscillator only affects the $b$ oscillator through the number $m_a$ and vice-versa. The eigenvectors and eigenvalues can readily be found via the using third-quantization and the procedure presented in Sec.~\ref{SM:Eigenvales_eigenvectors}. This reasoning applies to an arbitrary number of oscillators, as long as they each coupled to their own independent Markovian environment. 

Introducing dephasing still makes this problem solvable, and in fact proves to be trivial using our method. Indeed, not only is the weak $U(1)$ symmetry still present, but upon projecting onto the fixed $m_a$ subspace, it is found that the dephasing superoperator can be replaced by a number:
\begin{align}
    2\kappa_{\phi} 
    \mathcal{D}[\ha^\dagger \ha]\hrho
    =
    -\kappa_{\phi, a} 
    [
    \ha^\dagger \ha
    ,
    [
    \ha^\dagger \ha
    ,
    \hrho
    ]
    ]
    \to
    -\kappa_{\phi}
    m^2 \hrho_{m}.
\end{align}
Thus, dephasing only shifts the eigenvalues by $-\kappa_{\phi} m^2$, whereas the eigenvectors do not change.

%%%%%%%%%%%%%%%%%%%%%%%%%%%%%%%%%%%%%
%%%%%%%%%%%%%%%%%%%%%%%%%%%%%%%%%%%%%
\section{Eigenvectors and eigenvalues of the dissipative Ising model}
We now analytically find the eigenvalues and eigenvectors of the dissipative Ising model as written in Eq.~(12) in the main text. Introducing the superoperators 
	\begin{align}\label{eq:Def_L_R_sigma_m}
	\hsigmam_{L,j} \ket{\hrho} \equiv |\hat{\sigma}_j^- \hrho \rangle,
	\hspace{0.5cm}
	&\hsigmam_{R,j} \ket{\hrho} \equiv |\hrho \hat{\sigma}_j^- \rangle
	\\ \label{eq:Def_L_R_sigma_p}
	\hsigmap_{L,j} \ket{\hrho} \equiv |\hat{\sigma}_j^+ \hrho \rangle,
	\hspace{0.5cm}
	&\hsigmap_{R,j} \ket{\hrho} \equiv |\hrho  \hat{\sigma}_j^+\rangle
	\\ \label{eq:Def_L_R_sigma_z}
	\hsigmaz_{L,j} \ket{\hrho} \equiv |\hat{\sigma}_j^z \hrho \rangle,
	\hspace{0.5cm}
	&\hsigmaz_{R,j} \ket{\hrho} \equiv |\hrho  \hat{\sigma}_j^z\rangle
    \end{align}
    one can write the third-quantized form of the Lindbladian as
    \begin{align}\nonumber
        \hL
        &=
        -i\sum_{j<k}
        J_{jk}
        \left( 
        \hsigmaz_{L,j} \hsigmaz_{L,k}
        -
        \hsigmaz_{R,j} \hsigmaz_{R,k}
        \right)
        -i
        \sum_j
        h_j
        \left(
        \hsigmaz_{L,j}-\hsigmaz_{R,j}
        \right)
        +
        \sum_{j}
        \gamma_{-,j}
        \left(
        \hsigmam_{L,j} \hsigmap_{R,j}
        -
        \frac{1}{2}
        (
        \hsigmap_{L,j} \hsigmam_{L,j}
        +
        \hsigmam_{R,j} \hsigmap_{R,j}
        )
        \right)
        \\ \label{SM:L_Spin_Third_Quantized}
        &+
        \sum_{j}
        \gamma_{+,j}
        \left(
        \hsigmap_{L,j} \hsigmam_{R,j}
        -
        \frac{1}{2}
        (
        \hsigmam_{L,j} \hsigmap_{L,j}
        +
        \hsigmap_{R,j} \hsigmam_{R,j}
        )
        \right)
        +
        \sum_{j}
        \gamma_{\phi,j}
        \left(
        \hsigmaz_{L,j} \hsigmaz_{R,j}
        -
        \hone
        \right).
    \end{align}
 Note that our conventions imply $[\hsigmam_{L,j}, \hsigmap_{L,j}] = \hsigmaz_{L,j}$ and $[\hsigmam_{R,j}, \hsigmap_{R,j}] = -\hsigmaz_{R,j}$. Further, these superoperators inherent some of the properties of spin raising and lowering operators, such as $(\boldsymbol{\hat{\sigma}}^-_{\text{L/R}, j})^2 = (\boldsymbol{\hat{\sigma}}^+_{\text{L/R}, j})^2 = 0$ and $\{\boldsymbol{\hat{\sigma}}^-_{\text{L/R}, j}, \boldsymbol{\hat{\sigma}}^+_{\text{L/R}, j} \} = (\hsigmaz_{\rm {L/R}})^2 = \hone$.  In this model there are $N$ weak $U(1)$ symmetries, one for each spin. Since the master equation is invariant under $\hat{\sigma}^-_j \to e^{-i \theta_j} \hat{\sigma}^-_j$, the third-quantized version of the Lindbladian is unchanged when making the transformation $\boldsymbol{\hat{\sigma}}^-_{L/R, j} \to e^{-i \theta_j} \boldsymbol{\hat{\sigma}}^-_{L/R, j}$. Using the superoperator commutation relations, one can show that  the corresponding unitary operator is $e^{i \theta_j/2(\hsigmaz_{L,j}-\hsigmaz_{R,j})} \boldsymbol{\hat{\sigma}}^-_{L/R, j} e^{-i \theta_j/2(\hsigmaz_{L,j}-\hsigmaz_{R,j})} = e^{-i \theta_j} \hsigmam_{L/R, j}$. The Lindbladian therefore commutes with each generator, and thus can be written in block-diagonal form $\hL = \bigoplus_{\vec{M}} \hL_{\vec{M}}$ where $\vec{M} = \{m_1, \dots, m_N\}$ is a list indexing each block whose elements $m_j$ are eigenavlues of $(\hsigmaz_{L,j} - \hsigmaz_{R,j})/2$. The generators have two non-degenerate eigenvalues $m_j = 1$ and $m_j = -1$ with associated eigenoperator $\ket{\uparrow_j}\bra{\downarrow_j}$ and $\ket{\downarrow_j}\bra{\uparrow_j}$ respectively, in addition to a two-fold degenerate eigenvalue $m_j = 0$ with eigenoperators $\ket{\uparrow_j}\bra{\uparrow_j}$ and $\ket{\downarrow_j}\bra{\downarrow_j}$. The size of the block $\hL_{\vec{M}}$ is thus $2^{z(\vec{M})}$ where $z(\vec{M})$ is the number of zero eigenvalues in the list $\vec{M}$.\\
 
 Now we wish to use the main idea of this work presented in the main text: once we replace the generators by their eigenvalues in a given subspace, mean-field theory becomes exact. To that end, using the property $(\hsigmaz_{\rm {L/R}})^2 = \hone$, we can write 
\begin{align}\nonumber
        \hsigmaz_{L,j} \hsigmaz_{L,k}
        -
        \hsigmaz_{R,j} \hsigmaz_{R,k}
        &
        =
        \frac{1}{2}
        \left(
        (
        \hsigmaz_{L,j} + \hsigmaz_{L,k}
        )^2
        -
       (
        \hsigmaz_{R,j} + \hsigmaz_{R,k}
        )^2
        \right)
        \\
        &
        =
        \frac{1}{2}
        \left(
        \hsigmaz_{L,j}-\hsigmaz_{R,j}
        +
        \hsigmaz_{L,k}-\hsigmaz_{R,k}
        \right)
        \left(
        \hsigmaz_{L,j}+\hsigmaz_{R,j}
        +
        \hsigmaz_{L,k}+\hsigmaz_{R,k}
        \right).
\end{align}

Projecting $\hL$ onto the subspace indexed by $\Vec{M} = \{m_1, \dots, m_N\}$, we can then replace the generators by their eigenvalue $(\hsigmaz_{L,j}-\hsigmaz_{R,j})/2 \to m_j$. Within this subspace, the interaction then takes the form
\begin{align}\label{SM:Spin_Interaction}
    \sum_{j<k}
    J_{jk}
    (
    m_j+m_k
    )
    \left(
    \hsigmaz_{L,j}+\hsigmaz_{R,j}
    +
    \hsigmaz_{L,k}+\hsigmaz_{R,k}
    \right)
    =
    \sum_{j}
    J_j^{\rm eff}(\vec{M})
    \left(
    \hsigmaz_{L,j}+\hsigmaz_{R,j}
    \right)
\end{align}
where as in the main text we have defined 
\begin{align}
    J^{\rm eff}_j
    =
    \sum_{k \neq j}
    J_{jk} m_k.
\end{align}
 The exactness of mean-field theory in each block thus implies that we are now tasked with finding the eigenvalues of $N$ pseudo-Liouvillians which act independently on each spin:
\begin{align}
    \hL_{\vec{M}}
    =
    \sum_{j}
    \hL_j(\vec{M}).
\end{align}
The form of $\hL_j(\vec{M})$ follows immediately from Eq.~(\ref{SM:L_Spin_Third_Quantized}) and Eq.~(\ref{SM:Spin_Interaction}). As discussed in the main text, the fact that the $m_j = \pm 1$ eigenvalues are non-degenerate whereas the $m_j = 0$ are immediately implies that there are a set of frozen and active spins described by populations and coherence's respectively. Since the $m_j = 0$ eigenvalues are two-fold degenerate, we need another list $\vec{\chi} = \{ \chi_{j} \: | \: j \: \rm{frozen}\}$ of size $z(\vec{M})$ to label the eigenvectors and eigenvalues in each block where $\chi_{j} = \pm $. The right and left eigenvectors can thus be written as
\begin{align}
     | \hat{r}_{\vec{M}, \vec{\chi}} \rangle
     &=
     | \hat{r}_{\rm froz} \times \hat{r}_{\rm act} \rangle
     =
     |  \left(
        \prod_{j \: \rm{frozen}}
        \hat{\sigma}^{m_j}_j
        \right)
    \left(
    \prod_{j \: \rm active}
    \left(
    r_{\uparrow, \chi_j } \ket{\uparrow_j}\bra{\uparrow_j} 
    + 
    r_{\downarrow, \chi_j} \ket{\downarrow_j}\bra{\downarrow_j} 
    \right)
    \right)
    \rangle
    \\
    \langle \hat{l}_{\vec{M}, \vec{\chi}} |
     &=
     \langle \hat{l}_{\rm froz} \times \hat{l}_{\rm act} |
     =
     \langle  \left(
        \prod_{j \: \rm{frozen}}
        \hat{\sigma}^{m_j}_j
        \right)
    \left(
    \prod_{j \: \rm active}
    \left(
    l_{\uparrow, \chi_j } \ket{\uparrow_j}\bra{\uparrow_j} 
    + 
    l_{\downarrow, \chi_j} \ket{\downarrow_j}\bra{\downarrow_j} 
    \right)
    \right)
    |
\end{align}
where, for the sake of compactness, we have dropped the explicit $\vec{M}$ dependence of $r_{\uparrow/\downarrow, \chi_j}$ and $l_{\uparrow/\downarrow, \chi_j}$. The eigenvalues can then be written as
   \begin{align}
        \lambda_{\vec{M},\vec{\chi}}
        =
        \sum_{j}
        \lambda_j
        (
        \vec{M},
        \chi_j
        )
    \end{align}
For frozen spins, one does not need to solve an eigenvalue problem: to obtain $\lambda_j(\vec{M}, \chi_j)$ one must simply apply $\hL_j(\vec{M})$ to the coherences and obtain the eigenvalue Eq.~(16) in the main text. A simple calculation reveals that the eigenvectors and eigenvalues of the active spins satisfy:
    \begin{align}\label{SM:m_0_r_Eigenvalue}
        \begin{pmatrix}
         -2i J_{j}^{\rm eff}(\vec{M})-\gamma_{-,j}& \gamma_{+, j} \\
         \gamma_{-,j} & 2 i J_{j}^{\rm eff}(\vec{M})-\gamma_{+,j} 
        \end{pmatrix}
        \begin{pmatrix}
        r_{\uparrow, \chi_j}
        \\
        r_{\downarrow, \chi_j}
        \end{pmatrix}
        =
        \lambda_{j}(\vec{M},s_j)
        \begin{pmatrix}
        r_{\uparrow, \chi_j}(\vec{M})
        \\
        r_{\downarrow, \chi_j}(\vec{M})
        \end{pmatrix},
    \end{align}
    \begin{align}\label{SM:m_0_l_Eigenvalue}
    \begin{pmatrix}
     -2i J_{j}^{\rm eff}(\vec{M})-\gamma_{-,j}& \gamma_{-, j} \\
         \gamma_{+,j} & 2 i J_{j}^{\rm eff}(\vec{M})-\gamma_{+,j} 
        \end{pmatrix}
        \begin{pmatrix}
        l^*_{\uparrow, \chi_j}
        \\
        l^*_{\downarrow, \chi_j}
        \end{pmatrix}
        =
        \lambda_{j}(\vec{M},\chi_j)
        \begin{pmatrix}
        l^*_{\uparrow, \chi_j}
        \\
        l^*_{\downarrow, \chi_j}
        \end{pmatrix}.
    \end{align}
The above equations are readily solved, and we obtain
\begin{align}
    r_{\uparrow, \pm}
    &=
    \frac{
    \eta_j(\Gamma_j+\eta_j)
    }
    {
    \Gamma_j
    \left(2(\eta_j-i J^{\rm eff}_j))
    +
    \Gamma_j
    \mp 
    \sqrt{\Gamma_j^2-4 J^{\rm eff}_j(J^{\rm eff}_j+i \eta_j)}
    \right)
    }
    \\
    r_{\downarrow,  \pm}
    &=
    \frac{
    \eta_j(\Gamma_j-\eta_j)
    }
    {
    \Gamma_j
    \left(2(\eta_j-i J^{\rm eff}_j))
    -
    \Gamma_j
    \pm
    \sqrt{\Gamma_j^2-4 J^{\rm eff}_j(J^{\rm eff}_j+i \eta_j)}
    \right)
    }
\end{align}
\begin{align}
    l^*_{\uparrow, \pm}
    &
    =
    \pm
    \frac{\Gamma_j}{2\eta_j} 
    \frac{
    2(\eta_j-i J_j^{\rm eff})
    - \Gamma_j 
    \pm
    \sqrt{\Gamma_j^2-4 J^{\rm eff}_j(J^{\rm eff}_j+i \eta_j)}
    }
    {\sqrt{\Gamma_j^2-4 J^{\rm eff}_j(J^{\rm eff}_j+i \eta_j)}}
    \\
    l^*_{\downarrow, \pm}
    &
    =
    \pm
    \frac{\Gamma_j}{2\eta_i}
    \frac{
    2(\eta_j-i J_j^{\rm eff})
    +\Gamma_j 
    \mp
    \sqrt{\Gamma_j^2-4 J^{\rm eff}_j(J^{\rm eff}_j+i \eta_j)}
    }
    {\sqrt{\Gamma_j^2-4 J^{\rm eff}_j(J^{\rm eff}_j+i \eta_j)}}
\end{align}
where as in the main text $\Gamma_j = (\gamma_{+,j}+\gamma_{-,j})/2$, $\eta_j = (\gamma_{+,j}-\gamma_{-,j})/2$ The normalization has been chosen such that the biorthogonality condition
\begin{align}
    \langle \hat{l}_{\vec{M}',\vec{S}'} | \hat{r}_{\vec{M},\vec{S}} \rangle
    =
    \delta_{\vec{M} ,\vec{M}'}
    \delta_{\vec{S},\vec{S}'}
\end{align}
is satisfied. Further, we have also chosen the normalization such that the steady state has unit trace. The corresponding eigenvalues are then found to match those found in Eq.~(16) in the main text.

    %%%%%%%%%%%
	% FIG SM 1 Kappa
	%%%%%%%%%%%
	\begin{figure}[t]
	\centering
    \includegraphics[width=0.45 \textwidth]{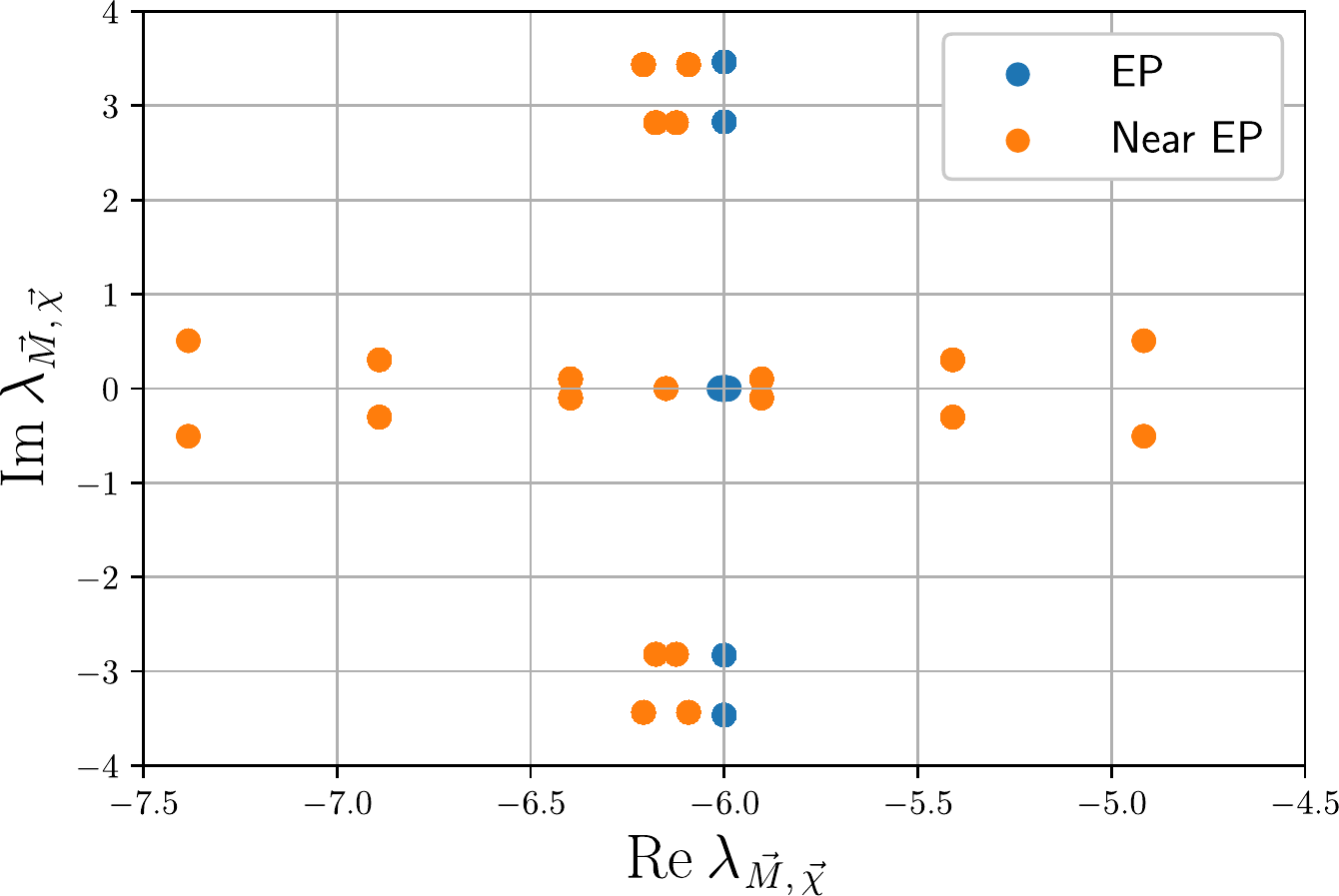}
	\caption{Select eigenvalues of the dissipative Ising Lindbladian as written in Eq.~(\ref{SM:L_Spin_Third_Quantized}) at an EP ($\gamma_{+, j} = \gamma_{-,j} = 1$ and $J = 0.5$) and near an EP ($\gamma_{+, j} = 1.05, \gamma_{-,j} = 1$ and $J = 0.5$). The perturbation $\delta \gamma_{+} = 0.05$ causes a much larger shift of the eigenvalues on the order of $\sqrt{\delta \gamma_+} = 0.22$.
	}
	\label{fig:EP}
	\end{figure}
    %%%%%%%%%%%%%%%%%%%%%%%%%%%%%%%%%%%%%
    %%%%%%%%%%%%%%%%%%%%%%%%%%%%%%%%%%%%%

We end this section by pointing out the existence of several exceptional point (EP) in this model. An EP occurs when both the eigenvalues and eigenvectors of a non-Hermitian matrix coalesce, thus rendering it undiagonalizable \cite{Heiss2012_EP_Review}. The order of the degeneracy is know as the order of the EP, and it is well-know that at an EP, the spectrum is exceptionally sensitive to perturbations $\epsilon$ of certain parameters, where the change of the eigenvalues scale like $(\epsilon)^{1/n}$  \cite{Heiss2012_EP_Review}. Since our problem reduces to a solving a set of $2 \times 2$ matrices, it is clear that we can have at most second-order EPs. It is easy to see that these necessarily occur when
\begin{align}
     \sqrt{\Gamma^2_j - 4 J_{j}^{\rm eff}(J_j^{\rm eff}+i \eta_j)} = 0
     \implies 
     \eta_j 
     = 
     (\gamma_{+, j}-\gamma_{-, j})/2 = 0
     \: \: \:
     \text{and}
     \: \: \:
     \Gamma_j = \pm 2 J_{j}^{\rm eff}.
\end{align}
To obtain an EP, the incoherent pumping and decay rate must therefore be the same. One way to obtain a large number of second-order EP's is then to make both the incoherent pumping and decay uniform $\gamma_{+,j} = \gamma_{-, j} = \Gamma_j = \gamma$  and set all the coherent interaction terms to be uniform as well $J_{j k } = J = \gamma/2$. Then $J_{j}^{\rm eff} = J \sum_{m \neq k} m_k$ and we obtain a set of EPs when $\sum_{m \neq k} m_k = \pm 1$. In Fig.~(\ref{fig:EP}) we plot the spectrum near and at an EP. We see that the small perturbation induces a large change in the eigenvalues.

%%%%%%%%%%%%%%%%%%%%%%%%%%%%%%%%%%%%%
%%%%%%%%%%%%%%%%%%%%%%%%%%%%%%%%%%%%%
%%%%%%%%%%%%%%%%%%%%%%%%%%%%%%%%%%%%%
%%%%%%%%%%%%%%%%%%%%%%%%%%%%%%%%%%%%%
\section{Decay of single-spin coherence in the dissipative Ising model}
In this section we compute the single-spin coherence $\langle \hat{\sigma}_j^+ (t) \rangle$ using the eigenvalue and eigenvector decomposition of the Lindbladian discussed in the previous section, assuming that all spins at time $t = 0 $ are initially in the $x$ direction. Using third-quantized notation, we have
\begin{align}\label{SM:Spin_Coherence}
    \langle \hat{\sigma}_j^+ (t) \rangle
    =
    \langle \hat{1} |
    \hsigmap_{L, j}
    e^{\hL t}
    |
    \hrho(0)
    \rangle 
    =
    \langle \hat{1} |
    \hsigmap_{L, j}
    \sum_{\vec{M}, \vec{\chi}}
    e^{\lambda_{\vec{M}, \vec{\chi}} t}
    | \hat{r}_{\vec{M}, \vec{\chi}} \rangle 
    \langle \hat{l}_{\vec{M}, \vec{\chi}} 
    |
    \hrho(0)
    \rangle 
    =
    \sum_{\vec{M}, \vec{\chi}}
    e^{\lambda_{\vec{M}, \vec{\chi}} t}
    \Tr(
    \hat{\sigma}^+_j \hat{r}_{\vec{M}, \vec{\chi}}
    )
    \Tr(
    \hat{l}_{\vec{M}, \vec{\chi}}^\dagger \hrho(0)
    )
\end{align}
Note that the block-diagonal structure greatly simplifies the summation above, as we are only probing a specific set of eigenmodes. More concretely, the only modes that contribute are those for which the only frozen spins is spin $j$, whereas all others are active:
\begin{align}
    \Tr(
    \hat{\sigma}^+_j \hat{r}_{\vec{M}, \vec{\chi}}
    )
    =
    \begin{cases}
    \prod_{k \neq j}
    (
    r_{\uparrow, \chi_k}
    +
    r_{\downarrow, \chi_k}
    )
    ,
    \hspace{0.5 cm}
    m_{j} = -1, \:  m_{k} = 0 \:  \forall \:  k \neq j
    \\
    0
    ,
    \hspace{0.5 cm}
    \text{otherwise}.
    \end{cases}
\end{align}
There are thus only $2^{N-1}$ out of the $4^N$ terms in Eq.~(\ref{SM:Spin_Coherence}) that do not vanish. The fact that the eigenvectors and initial state factorize over a product of single spins, in addition to the fact that the eigenvalues can be written as $ \lambda_{\vec{M},\vec{\chi}} = \sum_{j} \lambda_j ( \vec{M},\chi_j)$ directly leads the factorization of the end result
\begin{align}\nonumber
    \langle \hat{\sigma}_j^+ (t) \rangle
    &=
    \frac{e^{(2i h_j -\Gamma_j -2 \gamma_{\phi, j} ) t}}{2}
    \prod_{k \neq j}
    \left(
    \sum_{\chi_k = \pm }
    \frac{
    e^{\lambda_{k}(\vec{M}, \chi_k)t}
    }
    {2}
    \left[
    (
    r_{\uparrow, \chi_k}
    +
    r_{\downarrow, \chi_k}
    )
    (
    l^*_{\uparrow, \chi_k}
    +
    l^*_{\downarrow, \chi_k}
    )
    \right]
    \right)
    \\ \label{SM:Spin_Decoherence}
    &
    =
    \frac{e^{(2i h_j -\Gamma_j -2 \gamma_{\phi, j} ) t}}{2}
    \prod_{k \neq j}
    \left(
    e^{-\Gamma_k t}
    \left[
    \cosh(S_k)
    +
    \frac{\Gamma_k}{S_k}
     \sinh(S_k)
    \right]
    \right)
\end{align}
where $S_k = \sqrt{\Gamma^2_k-4 J^{\rm eff}_{k}(J^{\rm eff}_k + i \eta_k)}$ and $J_{k}^{\rm eff} = -J_{kj}$. This matches the result in Ref.~\cite{Ana_Maria_Diss_Ising_PRA_2013}, yet Eq.~(\ref{SM:Spin_Decoherence}) was obtained with only a few lines of algebra. Further, we note that in principle have access to the full many-body density matrix of the time-evolved initial state. 

\bibliography{SM_Weak_Symmetries}